\documentclass[11pt]{article}
\usepackage[left=1.5cm,right=1.5cm]{geometry}
\usepackage[numbers,square]{natbib}
\usepackage[T1]{fontenc}
\usepackage{tikz-cd}
\usepackage[british]{babel}
\usepackage{layout}
\usepackage{amsmath}
\usepackage{amssymb,amstext, amsthm, simplewick, amsfonts}
\usepackage{framed}
\usepackage{amsfonts}
\usepackage{color} 
\usepackage{braket}
\usepackage{multicol} 
\usepackage{epsfig}
\usepackage{cancel}
\usepackage{caption}
\usepackage{epsf}
\usepackage{feynmf} 
\usepackage{frontespizio}
\usepackage{rotating}
\usepackage{subfigure}
\usepackage{pstricks}
\usepackage{type1ec}
\usepackage{lettrine}
\usepackage{bbold}
\usepackage{calligra}
\usepackage{tikz}
\usepackage{float}
\usepackage{subfigure}
\usepackage{slashed}
\usepackage{mathrsfs}
\usepackage{curve2e}
\usepackage{setspace}
\usepackage{indentfirst}
\usepackage{emptypage}
\usepackage{relsize}
\usepackage{mathrsfs}
\usepackage{stackengine}
\usepackage{calc}
\usepackage{hyperref}
\hypersetup{
	colorlinks,
	citecolor=green,
	filecolor=black,
	linkcolor=blue,
	urlcolor=black
}

\newcommand{\3}[1]{C_{
		\ifthenelse{\equal{\ThreePt}{\empty}}{#1}{
			\ifthenelse{\equal{#1}{\empty}}{\ThreePt}{\ThreePt,#1}}}}
\newcommand{\redef}[1]{{C'}_{
		\ifthenelse{\equal{\ThreePt}{\empty}}{#1}{
			\ifthenelse{\equal{#1}{\empty}}{\ThreePt}{\ThreePt,#1}}}}
\newcommand{\ren}[1]{C_{
		\ifthenelse{\equal{\ThreePt}{\empty}}{#1}{
			\ifthenelse{\equal{#1}{\empty}}{\ThreePt}{\ThreePt,#1}}}}
\newcommand{\sd}[1]{D_{
		\ifthenelse{\equal{\ThreePt}{\empty}}{#1}{
			\ifthenelse{\equal{#1}{\empty}}{\ThreePt}{\ThreePt,#1}}}}

\numberwithin{equation}{section} 
\makeatletter
\@addtoreset{equation}{section}
\newcommand{\bea}{\begin{eqnarray}}
\newcommand{\eea}{\end{eqnarray}}
\makeatother
\newcommand{\beqa}{\begin{eqnarray}}
	\newcommand{\eeqa}{\end{eqnarray}}

\newcommand{\nn}{\nonumber}
\let\a=\alpha   \let\b=\beta      \let\d=\delta
         
\let\i=\iota      \let\l=\lambda  \let\m=\mu
\let\n=\nu           \let\p=\pi      \let\r=\rho
\let\s=\sigma

\newcommand{\bann}{\begin{eqnarray*}}
\newcommand{\eann}{\end{eqnarray*}}

\newcommand{\bmi}{\begin{minipage}}
\newcommand{\emi}{\end{minipage}}

\let\G=\Gamma  \let\D=\Delta   
           
    \let\Y=\Psi

\newcommand{\la}{\langle}
\newcommand{\ra}{\rangle}
\newcommand{\bs}[1]{\boldsymbol{#1}}

\newcommand{\be}{\begin{equation}}
	\newcommand{\ee}{\end{equation}}
\newcommand{\sdfrac}[2]{\mbox{\small$\displaystyle\frac{#1}{#2}$}}

\newcommand{\beq}{\begin{equation}}
	\newcommand{\eeq}{\end{equation}}
\newcommand{\figref}[1]{Fig.~\ref{#1}}			

\newcommand{\ThreePt}{\empty}
\usepackage{accents}
\makeatletter
\newcommand{\xLine}[2][]{\ext@arrow 0359\Rightarrowfill@{#1}{#2}}
\makeatother
\xdefinecolor{darkgreen}{RGB}{102, 204, 70}
\xdefinecolor{darkblue}{RGB}{0, 0, 153}
\usepackage{amssymb}
\usepackage{pifont}
\newcommand{\bes}{\begin{subequations}}
	\newcommand{\ees}{\end{subequations}}

\usepackage{tikz-feynman}
\usetikzlibrary{arrows}
\usetikzlibrary{snakes}
\usetikzlibrary{decorations,decorations.pathmorphing,decorations.markings}

\tikzset{graviton/.style={decorate, decoration={snake}, double}}
\tikzset{gluon/.style={decorate, decoration={coil, segment length=8, aspect=1.2, amplitude=3 }}}

\begin{document}
	\begin{center}
		\vspace{1.5cm}
	\begin{center}
	\vspace{1.5cm}
	{\Large \bf  A Dilaton Sum Rule for the Conformal Anomaly Form Factor\\}
{\Large \bf   in QCD at Order $\alpha_s$ \\}
	
	\vspace{1.4cm}

{\large \bf Claudio Corian\`o, Stefano Lionetti, Dario Melle and Leonardo Torcellini\\}

\vspace{0.3cm} 
	
	\vspace{0.3cm}
		\vspace{1cm}

{\it  Dipartimento di Matematica e Fisica, Universit\`{a} del Salento \\
and INFN Sezione di Lecce,Via Arnesano 73100 Lecce, Italy\\
National Center for HPC, Big Data and Quantum Computing\\}
{\it and CNR-nanotec, Lecce \\}

\vspace{0.5cm}

	\end{center}
	\begin{abstract}
We present an off-shell dispersive analysis of the graviton–gluon–gluon ($TJJ$) vertex, extending previous investigations carried out in both QED and QCD. Within the framework of a non-Abelian gauge theory, we extract the conformal anomaly form factor from the trace component of the correlator and demonstrate that it satisfies a one-loop sum rule, valid under the most general kinematic conditions. 
Analogously to the chiral and chiral-gravitational cases, a spectral flow emerges in which the exchanged intermediate state becomes localized at zero invariant mass along the graviton line as the quark mass approaches zero. The total integral of the spectral density precisely reproduces the anomaly. We examine how the behaviour of these spectral densities evolves as the system approaches the conformal limit with on-shell gluons. The perturbative analysis reveals that such sum rules are fundamental dynamical features of anomaly-induced interactions. In particular, the appearance or absence of associated dilaton poles is closely tied to whether the sum rule is saturated by a pole contribution or by a dispersive continuum. In the conformal, on-shell limit, the particle-pole interaction yields a nonlocal S-matrix element entirely supported on the light-cone.
\end{abstract}
	\end{center}
	\newpage

\section{Introduction}
In the context of gauge theory, the conformal anomaly, which characterizes the breaking of scale invariance, manifests through the quantum relation \cite{Collins:1976yq,Adler:1976zt,Chanowitz:1972vd}
\begin{equation}
\label{oper}
\eta_{\mu\nu}\langle T^{\mu\nu}\rangle = -\frac{\beta(g)}{2 g} \mathcal A(z),  \qquad \mathcal A(z)=F_{\mu\nu}^a F^{a \,\mu\nu},
\end{equation}  
where \(T^{\mu\nu}\) is the canonical stress-energy tensor, symmetrized via the Belinfante procedure, or alternatively, defined by embedding the action of the gauge theory in a gravitational background,
by differentiation of the corresponding partition function with respect to the metric $g_{\mu\nu}$. In a non-Abelian theory such as QCD, the trace anomaly relation \eqref{oper} can be derived in perturbation theory by an analysis of the diagram in Fig. \ref{expansion} if we consider only massless quarks in the virtual corrections. In the presence of fermion mass corrections, \eqref{oper} is modified, as we are going to describe below. In order to characterize this point, one defines the partition function
\begin{equation}
Z[g, A] = \int \, \mathcal{D}\psi \, \mathcal{D}\bar{\psi} \, e^{i S_{\text{QCD}}[A, \psi_i, \bar{\psi}, g]},
\end{equation}
with $S_{QCD}$ being the QCD action, with quarks $\psi_i$  $(i=1,\ldots n_f)$ and where $A^{\alpha a}$ is the gauge field of color index $a$. The symmetric stress energy tensor is defined as
\begin{equation}
T^{\mu\nu}(x) = \frac{2}{\sqrt{g(x)}} \frac{\delta S_{\text{QCD}}}{\delta g_{\mu\nu}(x)},
\end{equation}
and its quantum average in the gravitational background is derived as

\begin{equation}
\langle T^{\mu\nu}(x) \rangle = \frac{2}{\sqrt{g(x)}} \frac{\delta \ln Z[g]}{\delta g_{\mu\nu}(x)}.
\label{average}
\end{equation}

In \eqref{oper}, \(\beta(g)\) 
\beq
\beta(g)= {1 \over 3} \, \,  \, \, \frac{g_s^3}{16\pi^2} \, (- 11C_A + 2 n_f)
\label{betagg}
\eeq
denotes the beta function of the theory, with $C_A=N_c$ for a gauge theory with $SU(N_c)$ symmetry, 
 and $n_f$ fermions. The functional expansion of the quantum relation \eqref{average} in the background of an external gauge field, defines correlators of the form $TJ^n$, $(n=2,3,4)$, with multiple insertions of gauge currents $(J)$, 
the lowest nonzero contribution being the $TJJ$, with two gluon currents extracted by a functional differentiation of \eqref{average}.\\
In a non-Abelian gauge theory, identifying each individual term in the trace anomaly relation, in principle, requires computing correlators involving three and four external gluon currents. However, in practice, determining the coefficient of the anomaly—namely, the non-Abelian $\beta$-function—only necessitates evaluating the $TJJ$ correlator. Specifically, the diagrams shown in Fig. \ref{expansion} enable a direct computation of Eq. \eqref{oper} by tracing over the $\mu\nu$ indices of the following correlator
\beq
\label{eww}
\Gamma^{\mu\nu\alpha\beta \, a b}(z,x,y) \equiv \frac{\delta^2 Z}{\delta A_\alpha^a(x) \delta A_\beta^b(y)}\langle T^{\mu\nu}\rangle \Big|_{A=0},
\eeq
which lies at the heart of our analysis. As we will show, building upon results obtained in previous work, the evaluation of this correlator allows one to identify—across general non-Abelian theories, and in QCD in particular—a sum rule for a quantity we define as the conformal anomaly form factor. This quantity plays a role analogous to that of the  chiral anomaly form factors appearing in correlators of chiral currents. Much like the form factors arising in the $AVV$ (axial-vector/vector/vector) and $ATT$ (axial-vector/tensor/tensor) correlators analyzed in earlier studies \cite{Coriano:2025ceu}, the conformal anomaly form factor is extracted via a systematic decomposition of the correlator in Eq. \eqref{eww} into its independent tensor structures, extracting the trace and the traceless parts together with a longitudinal/transverse decomposition with respect to the external momenta.

 \subsection{Mass corrections} 

A key feature of conformal anomalies is the clear separation between contributions arising from explicit mass terms and those that are genuinely anomalous. These second breaking terms are classified as explicit. 
This separation can be directly confirmed within perturbation theory by analyzing fundamental correlators involving a single insertion of the stress-energy tensor, using dimensional regularization with minimal subtraction. Indeed, in the case of QCD, extra contributions to the trace anomaly are generated 
by the inclusion of fermion mass corrections in the diagrams of Fig. \ref{expansion}. In the Standard Model, for example, in the spontaneously broken phase, such extra contributions are clearly proportional to the vev of the Higgs field. In this case, by taking the trace of correlators involving the stress-energy tensor and vector currents, one derives an anomalous Ward identity of the form
\begin{equation}
\Gamma^{\alpha\beta a b}(z,x,y) 
\equiv \eta_{\mu\nu} \left\langle T^{\mu\nu}(z) J^{ a \alpha}(x) J^{ b \beta}(y) \right\rangle 
= \frac{\delta^2 \mathcal A(z)}{\delta A_{\alpha}^{ a }(x) \delta A_{\beta}^{ b}(y)} + \left\langle {T^\mu}_\mu(z) J^{a \alpha  }(x) J^{b\beta}(y) 
\right\rangle,
\label{traceid1}
\end{equation}
where $a,b$ denote color indices. 
For multiple Abelian gauge interactions in the external lines, the anomaly simplifies to
\begin{equation} \label{TraceAnomaly}
\mathcal A(z) = -\sum_i \frac{\beta_i}{2 g_i} F^{\alpha\beta {(i)}}(z) F^{(i)}_{\alpha\beta}(z),
\end{equation}
where $\beta_i$ are the (mass-independent) beta functions for the gauge couplings $g_i$. In conformal theories where $\beta_i = 0$, the anomaly vanishes. The corresponding $\beta_i$ are just proportional to the number of fermions and scalars propagating in the virtual corrections, and coupled to the different Abelian gauge fields $A_\mu^{(i)}$. We recall that the QCD action, as any non Abelian gauge theory at $d=4$, is conformally invariant 
at the level of its classical action for massless quarks, before any gauge fixing. \\ 
We recall that a more general expression of the conformal anomaly, which involves not only the gauge-invariant background term $F^2$ but also other geometric invariants such as the square of the Weyl tensor and the Gauss--Bonnet term, is encoded in the $TTT$ (three-graviton) vertex, explicitly derived at least in the Abelian case~\cite{Coriano:2017mux,Coriano:2018bsy}. The most general conformal solution of this correlator can be obtained directly from free field theory, provided one allows for arbitrary numbers of massless scalars ($n_s$), massless fermions ($n_f$), and spin-one gauge fields ($n_V$) circulating in the loop corrections. The conformal solutions obtained in~\cite{Bzowski:2018fql,Bzowski:2013sza,Osborn:1993cr} are fully consistent with this perspective. In fact, the general solution to the conformal constraints for this correlator can be parameterized in terms of these independent constants, which correspond to the multiplicities introduced as constants of integration in the conformal Ward identities.\\
The two terms on the right-hand side of Eq.~\eqref{traceid1} can be identified explicitly. The first term comes from functional derivatives of $\mathcal A$, the genuine anomaly contribution, while the second arises from the explicit insertion of $T^\mu_\mu$ into the correlator. Their difference captures the purely anomalous contribution and allows one to isolate $\mathcal A$ through
\beq
\mathcal A^{\alpha\beta a b}(z) = \Gamma^{\alpha\beta a b}(z,x,y) - \left\langle T^\mu_\mu(z) J^{ a \alpha}(x) 
J^{ b \beta}(y) \right\rangle.
\eeq
This distinction is essential for extracting the true quantum anomaly from the stress-energy tensor's trace.\\
Beyond the anomalous term, the full dilaton–gauge–gauge vertex, defined by coupling a scalar (dilaton) field $\rho$ to 
$\Gamma^{\a\b a b}$, includes additional structure, captured by two form factors $\Sigma^{\alpha\beta a b}$ and $\Delta^{\alpha\beta a b}$ in the Standard Model, associated with loop effects from virtual fermions, gauge bosons, and scalars (including Higgs and Goldstone modes). The full momentum-space expression of $\Gamma^{\alpha\beta ab}$ can be derived by embedding the Standard Model Lagrangian in curved spacetime, performing functional variations with respect to the external metric and gauge fields, and subsequently taking the flat-space limit~\cite{2013JHEP...06..077C}. One then obtains

\begin{equation}
\Gamma^{\alpha \beta a b}(q,p_1,p_2) = (2\pi)^4\, \delta^4(q - p_1 - p_2)\, \frac{i}{\Lambda_{cf}} 
\left( \mathcal A^{\alpha \beta a b}(p_1,p_2) + \Sigma^{\alpha \beta a b}(p_1,p_2) + \Delta^{\alpha \beta a b}(p_1,p_2)\right),
\label{sep}
\end{equation}
with
\begin{align}
\mathcal A^{\alpha \beta a b}(p_1,p_2) &= \int d^4 x\, d^4 y\, e^{i p_1 \cdot x + i p_2 \cdot y}\, 
\frac{\delta^2 \mathcal A(0)}{\delta A^{\alpha a}(x)\delta A^{\beta b}(y)}, \\
\Sigma^{\alpha \beta a b}(p_1,p_2) + \Delta^{\alpha \beta a b}(p_1,p_2) &= \int d^4 x\, d^4 y\, e^{i p_1 \cdot x + i p_2 \cdot y}\, 
\left\langle T^\mu_\mu(0) J^{a\alpha }(x) J^{b\beta}(y) \right\rangle.
\end{align}
$\Lambda_{cf}$ denotes a conformal breaking scale, around which the conformal anomaly interaction can be described by a local effective theory.  This defines the coupling of a possible dilaton field to the conformal anomaly. 
Detailed computations of such "$\rho FF$" vertex, with $F$ denoting the field strength of neutral gauge boson of the electroweak sector $\gamma, Z$, are discussed in \cite{2013JHEP...06..077C}, and serve to disentangle the genuine anomaly from model-dependent loop corrections. In \eqref{sep}, $\Sigma^{\alpha \beta a b}$ denote electroweak corrections which are 1-particle irreducible, while $\Delta^{\alpha \beta a b}$ contain external mixing terms with the Higgs field and can be found in \cite{2013JHEP...06..077C}. \\
Correlators of this type have been studied in QED \cite{Giannotti:2008cv,Armillis:2009pq} and in the broken electroweak sector of the Standard Model \cite{Coriano:2011zk}. From a phenomenological perspective, they are also relevant in condensed-matter contexts (for example, in experiments on topological materials), because Luttinger’s approach \cite{Luttinger:1964zz} relates thermal perturbations to fictitious gravitational fields, following the Tolman--Ehrenfest formulation \cite{Tolman:1930ona}. This correspondence enables analogue-gravity studies in condensed matter \cite{Chernodub:2021nff}.

\begin{figure}
    \centering
    \begin{tikzpicture}
\begin{feynman}
    \vertex (i1);
    \vertex[right=1cm of i1] (a1);
    \vertex[right=1cm of a1] (b1);
    \vertex[right=3cm of a1] (o1) ;
    \vertex[below=1cm of i1] (i2) ;
    \vertex[right=1cm of i2] (a2);
    \vertex[right=3cm of a2] (o2) ;

    \vertex[above=1cm of a1] (t2);
    \vertex[above=2cm of a2] (t3);
    
    \vertex[above=1.5cm of b1] (i3) {$T^{\mu\nu}$};
    
    \diagram* { 
    (i1)  -- [fermion] (a1) -- [fermion] (b1)   --[fermion] (o1),
    
  (i2)  -- [anti fermion] (a2)  --[anti fermion] (o2),

  (i3)  -- [graviton] (b1),
  (a1)-- [gluon] (a2)
  };

  \node[] at (-0.35,-0.45) {$\pi\ $};
\node[] at (4.46,-0.45) {$\pi\ $};

\filldraw[fill=black, fill opacity=0.1] (-0.42,-0.5) ellipse (.45cm and 1.3cm);
\filldraw[fill=black, fill opacity=0.1] (4.42 ,-0.5) ellipse (.45cm and 1.3cm);

\draw[double, double distance=1.5,line width=1.2, postaction={decorate, decoration={
			markings,
			mark=at position 0.8 with  {\arrow{ latex} } } } ] (-1.8,-0.5) -- node [above] {} (-0.9,-0.5);

\draw[double, double distance=1.5,line width=1.2, postaction={decorate, decoration={
			markings,
			mark=at position 0.8 with  {\arrow{ latex} } } } ] (4.9,-0.5) -- node [above] {} (5.85,-0.5);
\end{feynman}
\end{tikzpicture}
\hspace{1cm}
\begin{tikzpicture}
\begin{feynman}
    \vertex (i1);
    \vertex[right=1cm of i1] (a1);
    \vertex[right=3cm of a1] (o1) ;
    \vertex[below=1cm of i1] (i2) ;
    \vertex[right=3cm of i2] (a2);
    \vertex[right=1cm of a2] (o2) ;

    \vertex[above=1cm of a1] (t2);
    \vertex[above=2cm of a2] (t3);
    \vertex[blob] (t1) at (2,1.3) {{$TJJ$}};
    \vertex[above=1.5cm of t1] (i3) {$T^{\mu\nu}$};
    
    \diagram* { 
    (i1)  -- [fermion] (a1)   --[fermion] (o1),
    
  (i2)  -- [anti fermion] (a2)  --[anti fermion] (o2),

  (i3)  -- [graviton] (t1),
  (t1) -- [gluon] (a1) ,
  (t1) -- [gluon] (a2),
  };
  
 \node[] at (-0.35,-0.45) {$\pi\ $};
\node[] at (4.46,-0.45) {$\pi\ $};

\filldraw[fill=black, fill opacity=0.1] (-0.42,-0.5) ellipse (.45cm and 1.3cm);
\filldraw[fill=black, fill opacity=0.1] (4.42 ,-0.5) ellipse (.45cm and 1.3cm);

\draw[double, double distance=1.5,line width=1.2, postaction={decorate, decoration={
			markings,
			mark=at position 0.8 with  {\arrow{ latex} } } } ] (-1.8,-0.5) -- node [above] {} (-0.9,-0.5);

\draw[double, double distance=1.5,line width=1.2, postaction={decorate, decoration={
			markings,
			mark=at position 0.8 with  {\arrow{ latex} } } } ] (4.9,-0.5) -- node [above] {} (5.85,-0.5);
\end{feynman}
\end{tikzpicture}
    \caption{Leading (left) and NLO contributions to the GFF of the pion.}
    \label{fig:1}
\end{figure}
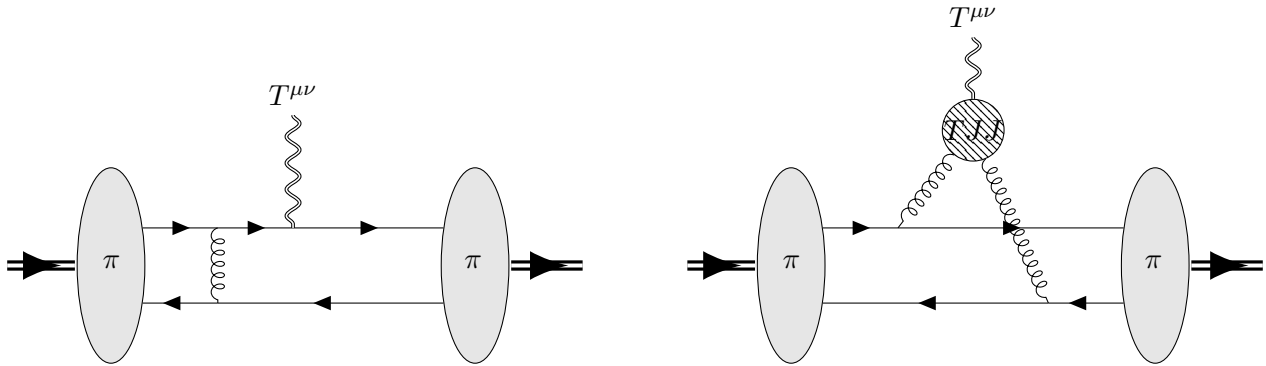

\subsection{The QCD case}
As mentioned above, in the perturbative analysis of a given anomaly vertex, the anomaly sector is characterized by the presence of a specific form factor, the anomaly form factor, which is constrained by a sum rule in its absorptive amplitude. This sum rule is associated with the emergence of an anomaly pole  - an interaction distinct from that of a standard particle pole - acting as an interpolating state in light-cone dominated kinematics.\\
This concept has been recently explored in \cite{Coriano:2025ceu}, particularly in the context of chiral and gravitational anomalies. The central aim of these studies has been to analyze the coupling and decoupling behavior of anomaly poles beyond light-cone kinematics, which defines the kinematical context in which their interaction is identical to that of an ordinary particle pole. There is, for a general kinematical setup, a clear distinction between these virtual states and conventional particle poles. To expose this behavior, explicit computations involving open indices in anomaly interactions are necessary. We will also proceed with a light-cone limit in order to illustrate 
which part of the $TJJ$ vertex dominates the interaction in this limit, as well as at small momentum transfers of the graviton 
line, showing its different behaviour compared to the case of chiral gauge and chiral gravitational interactions. In the present work, we investigate this interaction in detail for a non-Abelian theory. \\
In a related work we will show how this analysis applies to the case of the gravitational scattering of gauge fields.

The identification of the anomaly form factor within a non-Abelian gauge-fixed theory is subtle, as it involves disentangling contributions proportional to the gluon equations of motion from the genuine anomaly part. This requires a careful analysis of the interaction vertex both in the conformal limit and beyond, particularly for the $TJJ$ partonic correlator, 
which is the subject of our study, similarly to previous analysis of the chiral cases \cite{Coriano:2025ceu}. Analysis of these types of interactions in the parity-odd case - with the inclusion of anomalies - in CFT  can be found in 
\cite{Coriano:2023hts,Coriano:2023cvf,Coriano:2024ssu,Coriano:2023gxa,Coriano:2024nhv,Coriano:2024ive}.
The parity-even cases, involving the trace anomaly, had been studied in free field theory in QED in \cite{Giannotti:2008cv,Armillis:2009pq}, in QCD in \cite{Armillis:2010qk} and discussed in the general conformal approach in \cite{Bzowski:2013sza,Bzowski:2018fql}. Parity-odd trace anomaly have been addressed in \cite{Armillis:2010pa,Bonora:2014qla,Bastianelli:2019fot,Abdallah:2021eii,Larue:2023tmu,Larue:2023qxw}.\\
We recall that in Quantum Chromodynamics (QCD), the  perturbative $TJJ$ correlator plays a role in the computation of the gravitational form factors (GFFs) of the proton and the pion, albeit at subleading orders—$O(\alpha_s^2)$ and $O(\alpha_s^3)$ respectively—through standard QCD factorization of the GFFs. Indeed in  QCD, due to confinement, the quantum average in \eqref{oper} involves the computation of hadronic matrix elements, with profound implications for the analysis of the structure of hadrons at experimental level. For example, the  nucleon matrix element at zero momentum transfer $q=P'-P$, $\langle P |T^\alpha_\alpha | P \rangle$, directly relates to the nucleon mass squared. At nonzero momentum transfer, the matrix element of the uncontracted-stress energy tensor 
\beq
\langle P'| T^{\mu\nu} |P \rangle
\eeq
defining the GFF for a hadron, reveals a connection between such matrix element and the conformal anomaly \eqref{oper} for the strong interactions. \\
More generally, the interest in the GFF of the proton \cite{Tong:2022zax,Tanaka:2022wrr,Hatta:2018sqd,} is part of  efforts undertaken in the analysis of such hadronic matrix elements \cite{Tong:2022zax}, which are connected with the measurements of the Deeply Virtual Compton Scattering (DVCS) cross section and of generalized parton distributions \cite{Ji:1998pc,Radyushkin:1997ki} at JLab and at the 
Electron-ion collider (EIC) at Brookhaven. The link 
between GFFS and DVCS, in the proton case, is exemplified by the angular momentum sum rule for the analysis of the quark and gluon spin of the proton \cite{Ji:1996ek}. \\
Understanding the origin of the nucleon mass has become a central objective for the future EIC \cite{Abir:2023fpo}, together with a precise determination of the quark and gluon contributions to hadron spin. Once viewed as a largely conceptual issue with limited phenomenological impact, this decomposition has now become experimentally and computationally relevant. In our analysis, we will closely track the interaction terms associated with the quark and gluon contributions, as they exhibit distinct behavior with respect to conformal symmetry and to fermion–mass effects.

\subsection{The partonic $TJJ$ and the mass corrections}
In the case of a gauge theory such as QCD, as discussed in \cite{Coriano:2024qbr}, the operatorial relation \eqref{oper} emerges at leading order in the strong coupling \(g_s\) via a similar partonic \(TJJ\) three-point function, where \(T\) couples to virtual quarks and to the gluon currents $J^{\mu a}$, using ordinary factorization for exclusive processes. Due to the conformal invariance of the classical QCD action for massless quarks, the partonic vertex 
\beq
\Gamma^{\mu\nu\alpha\beta\, a b}\equiv \langle T^{\mu\nu}(q)J^{\alpha a}(p_1) J^{\beta b}(p_2)\rangle 
\eeq
can be investigated using approaches developed from CFT in momentum space, as shown for parity-even \cite{Bzowski:2013sza} and parity-odd \cite{Coriano:2023hts,Coriano:2023cvf,Coriano:2023gxa} tensor correlators. The connection between the partonic and the hadronic result can be clearly identified within the ordinary factorization picture of exclusive processes, in terms of hadronic waves functions and of hard 
scattering contributions which are affected by the partonic $TJJ$ vertex \cite{Coriano:2024qbr}. \\
For example, 
in the analysis of the gravitational form factor of the pion, the partonic $TJJ$ appears at $O(\alpha_s^2)$, 
with the $O(\alpha_s)$ contributions characterising the direct insertion of the QCD stress energy tensor, in its  quarks and gluons contributions, on the hard scattering.  \\
Our analysis, in this work, will be focused on the perturbative $TJJ$ correlator and , in particular, on its conformal anomaly form factor. Its absorptive part is characterised by the presence of a sum rule, as pointed out in previous works \cite{Horejsi:1997yn,Giannotti:2008cv} in the Abelian case, leaving a complete analysis of these features, at hadron level, to a future study. In the conformal limit, the trace anomaly manifests itself as a massless \(1/q^{2}\) pole in such form factor. This structure is isolated by employing a momentum–space, CFT-inspired tensor decomposition of the three-point vertex \cite{Bzowski:2013sza}. In this limit, the dispersive sum rule is trivially saturated by the residue of the pole. To assess the constraint beyond conformality, we move away from the conformal limit and include finite–fermion–mass corrections. In this broader framework, the classical scale invariance of QCD is explicitly broken by terms proportional to the fermion mass, so that \eqref{oper} is modified as

\beq
\eta_{\mu\nu}\langle T^{ \mu\nu}\rangle =m \langle \bar{\psi}\psi \rangle +\frac{ {\beta(g)}}{2 g}F^{\mu\nu a}F_{\mu\nu }^a, \label{qg}
\eeq
which includes contributions from the quark mass, in agreement with \eqref{traceid1}. 
The relation discussed above holds at both the partonic and hadronic levels. At leading order (tree level), the partonic \(TJJ\) correlator is partly fixed by conformal symmetry (in the quark sector) and furnishes a free-field realization of the QCD anomaly equation. Analyses of (chiral) anomalies at the partonic level have been revisited in several recent works \cite{Tarasov:2025mvn,Tarasov:2021yll,Bhattacharya:2023wvy,Bhattacharya:2022xxw,Castelli:2024eza}; the present study adds to this literature.

\subsection{Content of this work}
In this work, we first review the CFT-based approach to determining this correlator. We then systematically characterize the interaction at the conformal point, where QCD allows a direct connection to general CFT results in the conformal limit. Beyond this limit, we extend our analysis by incorporating quark mass corrections, which are essential in orde to proceed with an explicit analysis of the sum rule, once the anomaly form factor is correctly identified. \\
We explore the nature of the particle and anomaly poles that interpolate with this vertex in both cases, showing that the anomaly form factor of this correlator satisfies a sum rule at \(\mathcal{O}(\alpha_s)\). The proof of this sum rule, as well as the characteristic coupling/decoupling behavior of the anomaly pole in the correlator, is established through a direct analysis of the spectral density. This approach follows a strategy similar to that of \cite{Coriano:2025ceu}, which investigated the chiral and gravitational anomaly vertices in QCD at the same perturbative order. We demonstrate that the emergence of a dilaton-like anomaly pole in this interaction is directly linked to the sum rule governing its spectral density. \\
Building on previous analyses of the chiral and gravitational anomaly form factors, we examine the relationship between the conformal anomaly (dilaton) pole and the corresponding particle pole that interpolates in the interaction. 
In both cases, the coupling of these intermediate states is obtained by computing the corresponding residue, as detailed in Section 5. The spectral analysis of the conformal anomaly form factor follows the methodology outlined in \cite{Coriano:2025ceu} for the chiral and gravitational anomaly cases, with necessary modifications to account for the distinct nature of the correlator. Also in this case, as in previous studies, the anomaly pole interaction is defined on the light-cone, where the gluons are on-shell and the quarks are massless, but with some differences that we are going 
to describe in the final section of this work, before our conclusions.

\section{The partonic correlator in the conformal limit }
As previously noted, our entire analysis is conducted at the partonic level. 
QCD is not a conformal theory, but at tree level the Lagrangian is conformally invariant for vanishing quark masses. The free field theory realization of the $TJJ$ in QCD is indeed conformal, except for the corrections induced by renormalization of the corresponding correlator, which at order $g_s^2$  is simply obtained by a simple counterterm action shown in \eqref{cct}. This action is naturally coupled to gravity and the counterterm vertex generated from it, renormalizes the conformal Ward identities satisfied by the correlators. These become anomalous. As mentioned, 
our goal, from the open index structure of the correlator, is to isolate the specific form factor appearing in the trace part of the 
correlator, for which we verify the presence of a sum rule in the non-Abelian case, extending previous analysis. \\ 
The identification of such (conformal anomaly) form factor is achieved by combining CFT techniques on the tensorial decompositon of such correlators \cite{Bzowski:2013sza, Coriano:2013jba} with a perturbative analysis carried out away from the conformal point, by allowing a quark mass 
$m$ in the virtual corrections. The decomposition of the $TJJ$ in a non-Abelian gauge theory is significantly more intricate than in the standard CFT approach in momentum space, owing to the presence of a gauge-fixing sector that explicitly breaks conformal symmetry at one loop  \cite{Coriano:2024qbr}.\\
This complication does not arise in QED, where the conformal solution of the $TJJ$ correlator for Abelian currents precisely matches the general result obtained by solving the conformal Ward identities using momentum-space CFT methods ($\mathrm{CFT}_p$) \cite{Bzowski:2013sza, Coriano:2018bbe}.\\
The strategy adopted in \cite{Coriano:2024qbr} for the analysis of the $TJJ$ in QCD extends the decomposition method introduced in \cite{Bzowski:2013sza} to the non-Abelian case, by isolating the quark and gluon sector contributions within this correlator. This decomposition expresses the correlator in terms of transverse-traceless, longitudinal, and trace form factors. However, in QCD, the resulting expansion deviates from the conformal non-Abelian structure discussed in \cite{Bzowski:2013sza}, due to the appearance of Slavnov–Taylor identities, which supplant the ordinary Ward identities imposed in the Abelian case \cite{Armillis:2010qk} or in the absence of a gauge fixing sector in the action.
The form factors introduced in the tensorial decomposition allow to solve for the conformal constraints in this modified setup, 
in the presence of explicit breakings, just by resorting to the perturbative analysis in free field theory.  
We are going to discuss this decomposition, reviewing the basic results of \cite{Coriano:2025ceu} before moving to the spectral analysis of the conformal anomaly form factor of the $TJJ$. 
 
\subsection{The tensorial decomposition of the correlator}
In $CFT$ in momentum space, $CFT_p$, correlation functions are subject to the constraints of conformal Ward identities (CWIs), which serve as tensorial conditions in momentum space. In the quark sector of the correlator, the conformal constraints, for massless quarks, are similar to those valid in QED, with the inclusion of $SU(3)$ color indices $a_i$. In the general analysis of the correlator it is convenient, for simplicity, to adopt a symmetric notation and define all the momenta to be incoming, with $p_1+p_2+p_3=0$, where $p_1\equiv q$ is the momentum of the stress energy tensor expernal (graviton) with invariant mass $s\equiv p_1^2$. Similarly we will also using $s_1$ and $s_2$ to denote the virtualities of the two external gluons $(s_1\equiv p_1^2, s_2\equiv p_2^2)$.

The special conformal Ward identities  for a generic $TJ^n$ correlator 
take the form
\begin{equation}
	\begin{aligned}
			0=&\mathcal{K}^k\langle T^{\mu\nu}(q)J^{\mu_1 a_1}(p_1)\dots J^{\mu_{n-1} a_{n-1}}(p_{n-1})\rangle_q \equiv\\&\sum_{j=1}^{n-1}\left(p_j^\kappa \frac{\partial^2}{\partial {p_j}_\alpha \partial p_j^\alpha}+2\left(\Delta_j-d\right) \frac{\partial}{\partial p_j^\kappa}-2 p_j^\alpha \frac{\partial^2}{\partial p_j^\kappa \partial p_j^\alpha}\right)\langle T^{\mu\nu}(q)J^{\mu_1 a_1}(p_1)\dots J^{\mu_{n-1} a_{n-1}}(p_{n-1})\rangle_q\\&
		+2 \sum_{j=1}^{n-1} \left[\delta^{\kappa \mu_j} \frac{\partial}{\partial p_j^{\alpha_{j}}}-\delta_{\alpha_{j}}^\kappa \frac{\partial}{\partial {p_j}_{\mu_j}}\right]\langle T^{\mu\nu}(q)J^{\mu_1 a_1}(p_1)\dots J^{\alpha_j a_j}(p_j)\dots J^{\mu_{n-1} a_{n-1}}(p_{n-1})\rangle_q
	\end{aligned}
\end{equation}
while the dilatation WIs are given by
\begin{equation}
	0=\left[\sum_{j=1}^n \Delta_j-(n-1) d-\sum_{j=1}^{n-1} p_j^\lambda \frac{\partial}{\partial p_j^\lambda}\right]\langle T^{\mu\nu}(q)J^{\mu_1 a_1}(p_1)\dots J^{\mu_{n-1} a_{n-1}}(p_{n-1})\rangle_q,
\end{equation}
where both the stress energy tensor $T^{\mu\nu}$ and the gauge currents $ J^{a_i \, \mu_i}$ are decomposed into longitudinal-trace $(loc)$ and transverse-traceless components
 \cite{Bzowski:2013sza} 
\begin{align}
    T^{\mu_i\nu_i}(p_i) &\equiv t^{\mu_i\nu_i}(p_i) + t_{loc}^{\mu_i\nu_i}(p_i),\label{decT} \\
    J^{a_i \, \mu_i}(p_i) &\equiv j^{a_i \, \mu_i}(p_i) + j_{loc}^{a_i \, \mu_i}(p_i),\label{decJ}
\end{align}
where
\begin{align}
    t^{\mu_i\nu_i}(p_i) &= \Pi^{\mu_i\nu_i}_{\alpha_i\beta_i}(p_i) \,T^{\alpha_i \beta_i}(p_i), 
    && t_{loc}^{\mu_i\nu_i}(p_i) = \Sigma^{\mu_i\nu_i}_{\alpha_i\beta_i}(p) \,T^{\alpha_i \beta_i}(p_i), \label{loct} \\
    j^{a_i \, \mu_i}(p_i) &= \pi^{\mu_i}_{\alpha_i}(p_i) \,J^{a_i \, \alpha_i }(p_i), 
    && j_{loc}^{a_i \, \mu_i}(p_i) = \frac{p_i^{\mu_i} \, p_{i\,\alpha_i}}{p_i^2} \,J^{a_i \, \alpha_i}(p_i).
\end{align}
 The decomposition is performed using the transverse-traceless $(\Pi)$, transverse $(\pi)$, and longitudinal $(\Sigma)$ projectors, defined as
\begin{align}
    \pi^{\mu}_{\alpha} &= \delta^{\mu}_{\alpha} - \frac{p^{\mu} p_{\alpha}}{p^2}, \label{prozero} \\
    \Pi^{\mu \nu}_{\alpha \beta} &= \frac{1}{2} \left( \pi^{\mu}_{\alpha} \pi^{\nu}_{\beta} + \pi^{\mu}_{\beta} \pi^{\nu}_{\alpha} \right) 
    - \frac{1}{d - 1} \pi^{\mu \nu}\pi_{\alpha \beta}, \label{TTproj} \\
    \Sigma^{\mu_i\nu_i}_{\alpha_i\beta_i} &= \frac{p_{i\,\beta_i}}{p_i^2} 
    \Bigg[ 2\delta^{(\nu_i}_{\alpha_i)} p_i^{\mu_i)} - \frac{p_{i\alpha_i}}{(d-1)} 
    \left( \delta^{\mu_i\nu_i} + (d-2) \frac{p_i^{\mu_i} p_i^{\nu_i}}{p_i^2} \right) \Bigg] 
    + \frac{\pi^{\mu_i\nu_i}(p_i)}{(d-1)} \delta_{\alpha_i\beta_i}. \label{Lproj}
\end{align}
Additionally, we introduce
\begin{equation} \label{a:T}
    \mathscr{T}_{\mu\nu \alpha} (\bs{p}) = \frac{1}{p^2} \left[ 2 p_{(\mu} \delta_{\nu)\alpha} - \frac{p_\alpha}{d-1} \left( \delta_{\mu\nu} + (d-2) \frac{p_\mu p_\nu}{p^2} \right) \right].
\end{equation}
and the decomposition identity
\beq
    \delta_{\mu(\alpha}\delta_{\beta)\nu} = \Pi_{\mu\nu\alpha\beta}(\bs{p}) + \mathscr{T}_{\mu\nu (\alpha}(\bs{p})\,p_{\beta)} + \frac{1}{d-1}\pi_{\mu\nu}(\bs{p})\delta_{\alpha\beta}.
\eeq
As already pointed out, in the gluon sector the conformal equations are not valid any longer, due to the presence of the gauge-fixing. However, in \cite{Coriano:2024qbr} the approach 
has been to maintain the same tensorial sector decomposition used in $CFT_p$, which is valid to all orders in perturbation theory, and carefully identify all the extra contributions connected with the presence of virtual gluons in the perturbative expansion. The contributions from both sectors are shown in Fig. 1. \\ 
Applying this framework to the $TJJ$ correlator, we distinguish two fundamental components: the \textit{transverse-traceless} part and the \textit{semi-local} part, denoted by the subscript \( loc \). These components are determined by the transverse and trace Ward identities. Using the previously defined projectors, we decompose the full 3-point function as

\begin{align}
    \label{dec1}
    \braket{T^{\m \n}\,J^{ \, \a a}\,J^{\b \, b}} &= 
    \braket{t^{\m \n }\,j^{\a \, a}\,j^{\b \, b}}
    + \braket{T^{\m \n}\,J^{\a \, a}\,j_{loc}^{\b \, b}}
    + \braket{T^{\m \n}\,j_{loc}^{\a \, a}\,J^{\b \, b}} 
    + \braket{t_{loc}^{\m \n}\,J^{\a \, a}\,J^{\b \, b}} \notag \\
    &\quad - \braket{T^{\m \n }\,j_{loc}^{\a \, a}\,j_{loc}^{\b \, b}}
    - \braket{t_{loc}^{\m \n}\,j_{loc}^{\a \, a}\,J^{\b \, b}}
    - \braket{t_{loc}^{\m \n}\,J^{\a \, a}\,j_{loc}^{\b \, b}}
    + \braket{t_{loc}^{\m \n }\,j_{loc}^{\a \, a}\,j_{loc}^{\b \, b}}.
\end{align}
Using the projectors $\Pi$ and $\pi$ one can write the most general form of the transverse-traceless part as
	\begin{equation}
		{\braket{t^{\m \n}(p_1)\,j^{\a \, a}(p_2)\,j^{\b \, b }(p_3)}} =\Pi^{\m \n}_{\m_1 \n_1}(p_1)\pi^{\a}_{\a_1 }(p_2)\pi^{\b}_{\b_1}(p_3)\,\,X^{ \, \m_1 \n_1 \a_1 \b_1 \, a b},
	\end{equation}
	where $X^{a b \, \m_1 \n_1 \a_1 \b_1}_q$ is a general tensor of rank four built out of the metric and momenta. We can enumerate all possible tensors that can appear in $X^{a b \, \m_1 \n_1 \a_1 \b_1}$ preserving the symmetry of the 
	correlator, that in the quark case are represented by the functions  $A_{i}^{(q) a b}$ taking the form
	\begin{align}
		\langle t^{\mu \nu }(p_1)j^{a \, \a}(p_2)j^{b \, \b}(p_3)\rangle_q & =
		{\Pi_1}^{\mu \nu}_{\m_1 \n_1}{\pi_2}^{\a}_{\a_1}{\pi_3}^{\b}_{\b_1}
		\left( A_{1}^{(q) a b} \ p_2^{\m_1 }p_2^{\n_1}p_3^{\a_1}p_1^{\b_1} + 
		A_{2}^{(q) a b}\ \delta^{\a_1 \b_1} p_2^{\m_1}p_2^{\n_1} + 
		A_{3}^{(q) a b}\ \delta^{\m_1\a_1}p_2^{\n_1}p_1^{\b_1}\right. \notag\\
		& \left. + 
		A_{3}^{(q) a b}(p_2\leftrightarrow p_3)\delta^{\m_1\b_1}p_2^{\n_1}p_3^{\a_1}
		+ A_{4}^{(q) a b}\  \delta^{\m_1\b_1}\delta^{\a_1\n_1}\right)\label{DecompTJJx}.
	\end{align}
A similar expression for the transverse-traceless sector is generated by the virtual gluons 
holds, with form factors  $A_{i}^{(g) a b}$. The expression of the form factors identified by the expansion can be found in \cite{Coriano:2024qbr}.

\subsection {The reconstruction of the quark and gluon sectors in QCD}
The perturbative analysis requires the derivation of the QCD stress-energy tensor. 
It is obtained by adding the components of the field strength $(fs)$, fermionic, gauge fixing 
$(g. f)$ and ghost $(gh)$ sectors of the Lagrangian  
\bea
T_{\mu\nu} = T^{f.s.}_{\mu\nu} + T^{ferm.}_{\mu\nu} + T^{g.fix.}_{\mu\nu} + T^{ghost}_{\mu\nu},
\eea
with
\beq
T^{f.s.}_{\mu\nu}
 = 
-\eta_{\mu\nu}\frac{1}{4}F^a_{\r\s}F^{a\,\r\s}  + F^a_{\mu\r}F^{a\,\r}_\nu  
\eeq

\bea
T^{g.f.}_{\mu\nu} &=& -{1 \over \xi}\left[A_\nu^a \partial_\mu(\partial \cdot A^a) +A_\mu^a \partial_\nu(\partial \cdot A^a)\right] +{1 \over \xi}g_{\mu \nu}
\left[- \frac{1}{2} (\partial \cdot A)^2 + \partial^\rho (A_\rho^a \partial \cdot A^a)\right], \\
T^{gh}_{\mu\nu} &=&- \partial_{\mu}\bar c^a D^{ab}_{\nu} c^b - \partial_{\nu}\bar c^a D^{ab}_{\mu}c^b + g_{\mu\nu} \partial^{\rho}\bar c^a D^{ab}_{\rho}c^b.
\label{gfghost}
\eea
After the inclusion of the fermion sector, its final expression takes the form
\bea
T_{\mu \nu} &=& - g_{\mu \nu} {\mathcal L}_{QCD}
-F_{\mu \rho}^a F_\nu^{a \rho} -{\frac{1} {\xi}}g_{\mu \nu}
\partial^\rho (A_\rho^a \partial^\sigma A_\sigma^a) +{\frac{1}{\xi}}(A_\nu^a \partial_\mu(\partial^\sigma A^a_\sigma)
  +A_\mu^a \partial_\nu(\partial^\sigma A_\sigma^a))
\nonumber\\
&+& {\frac{i}{4}} \Big[
  \overline \psi \gamma_\mu (\overrightarrow \partial_\nu
-i g T^a A_\nu^a)\psi  -\overline \psi (\overleftarrow \partial_\nu
+i g T^a A_\nu^a)\gamma_\mu\psi
 +\overline \psi \gamma_\nu (\overrightarrow \partial_\mu
-i g T^a A_\mu^a)\psi
\nonumber\\
&-& \overline \psi (\overleftarrow \partial_\mu
+i g T^a A_\mu^a)\gamma_\nu\psi \Big] +\partial_\mu \overline
c^a (\partial_\nu c^a -g f^{abc} A_\nu^c  c^b)
+\partial_\nu \overline c^a (\partial_\mu c^a -g f^{abc}
A_\mu^c c^b).
\label{EMT}
\eea
 $\xi$ is an arbitrary gauge-fixing parameter and 
($c,\bar{c}$) are the ghost/antighost fields.
 Note that the stress energy tensor is normalized in the non-Abelian case in agreement with \cite{Armillis:2010qk}.  
 The derivation is performed by embedding the gauge-fixed QCD action into a gravitational background as described in 
 \cite{Coriano:2011zk} for the Standard Model with 
 \beq
 T^{\mu\nu }_{QCD} \equiv \frac{2}{\sqrt{-g}}\frac{\delta\mathcal{S}_{QCD}}{\delta g_{\mu\nu}}.
\eeq

\begin{figure}[t]
\begin{center}
\includegraphics[scale=0.10]{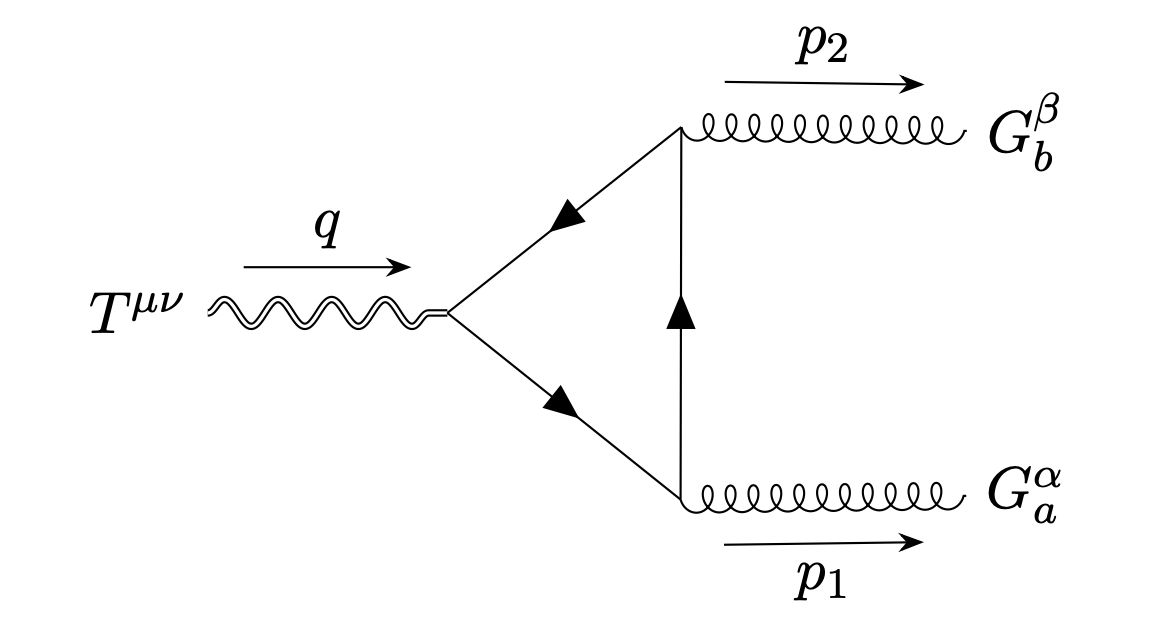}
\includegraphics[scale=0.10]{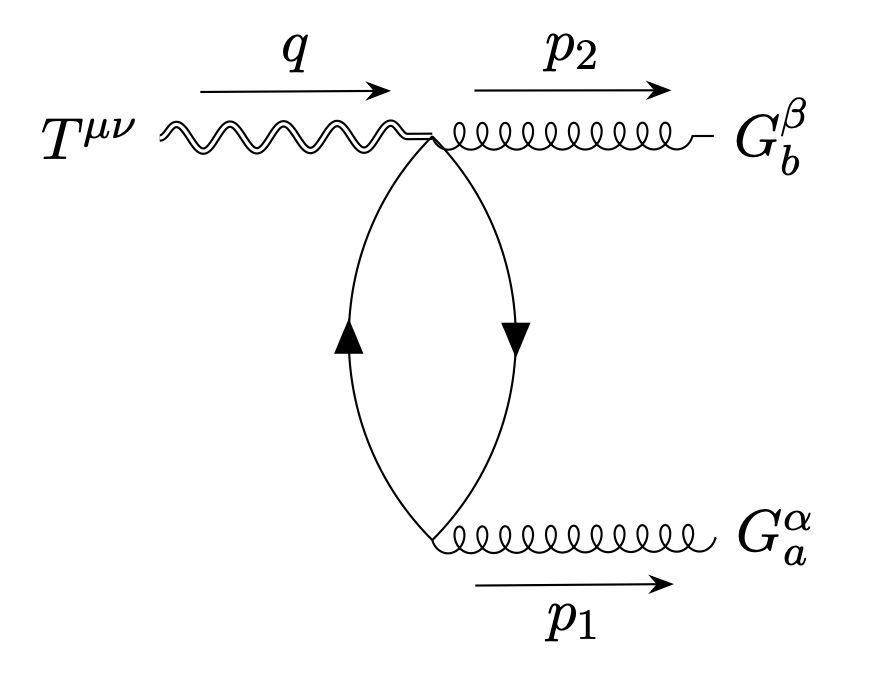}
\includegraphics[scale=0.10]{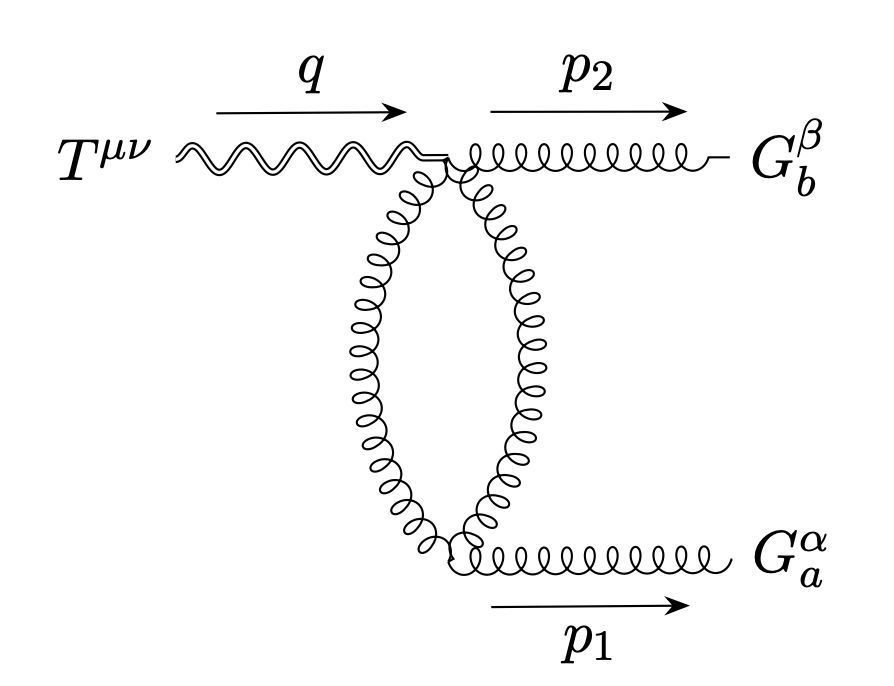}
\includegraphics[scale=0.10]{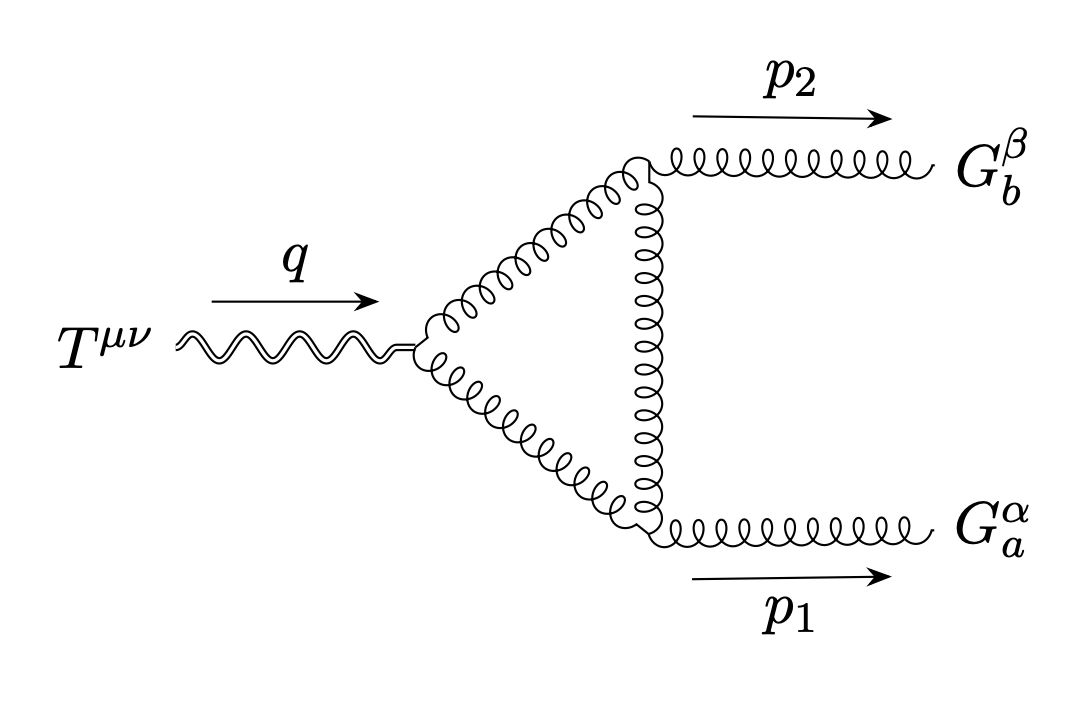}
\includegraphics[scale=0.10]{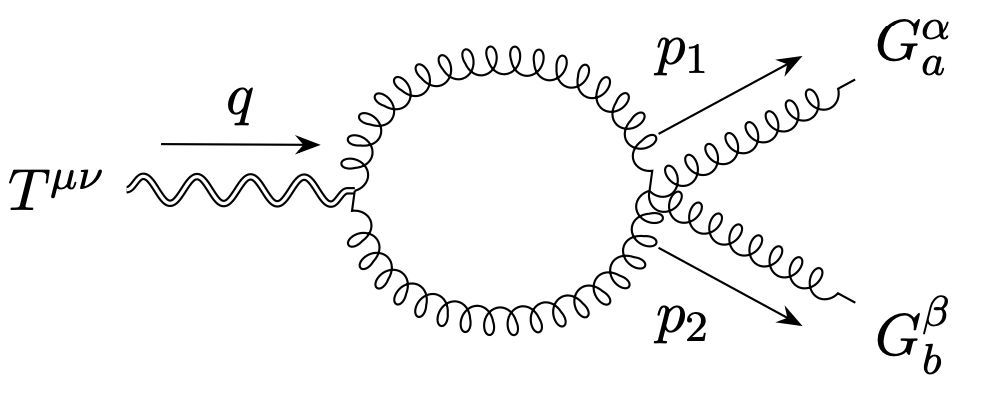}
\includegraphics[scale=0.10]{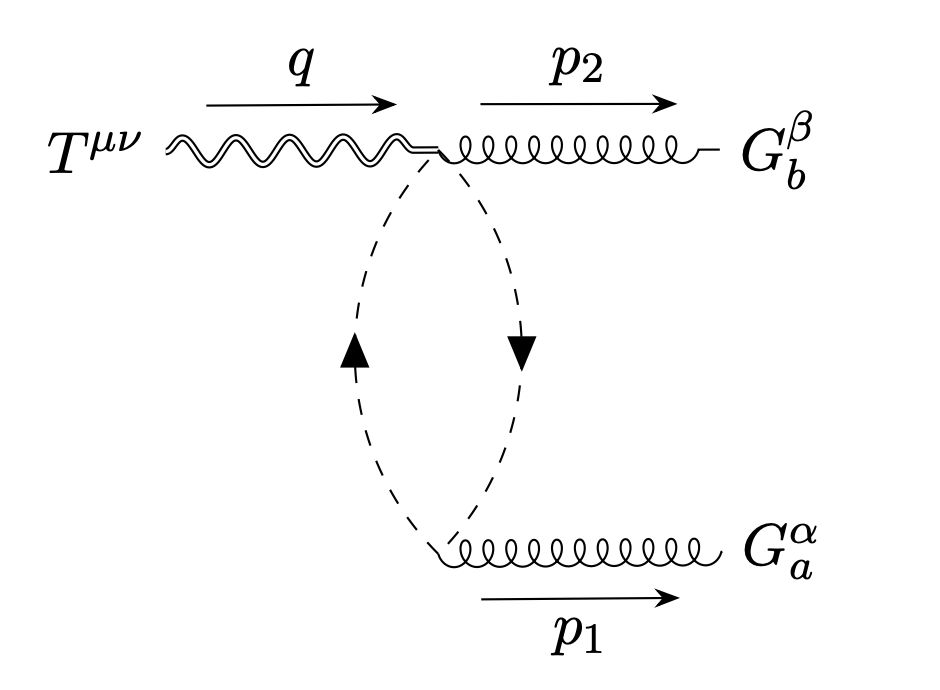}
\includegraphics[scale=0.10]{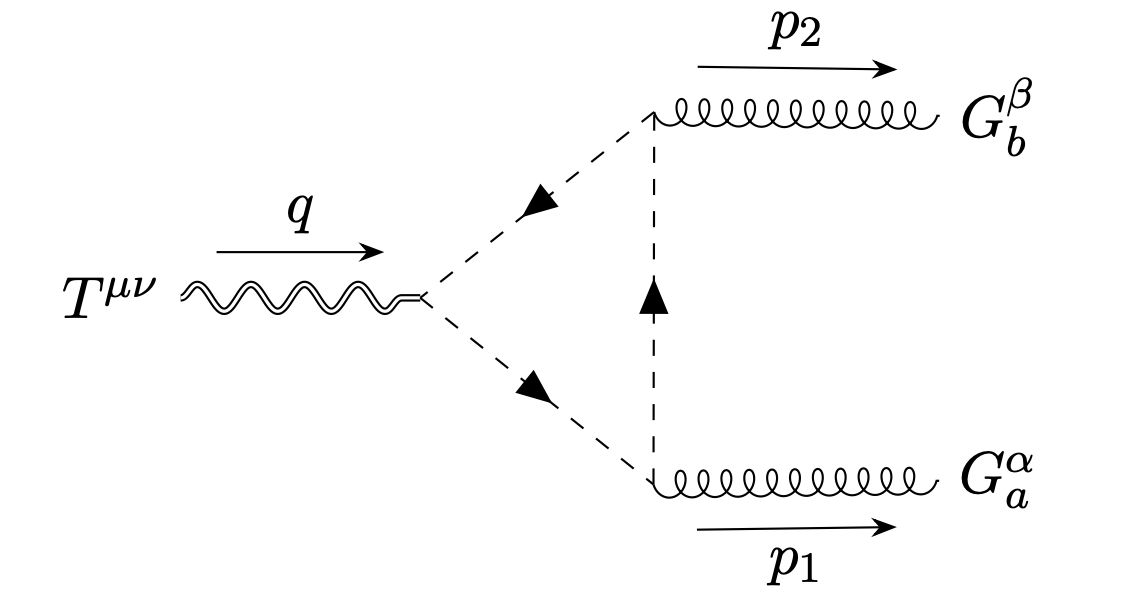}
\caption{List of the perturbative quark, gluon, and ghost (dashed lines) sectors in the non-Abelian $TJJ$. }. 
\label{expansion}
\end{center}
\end{figure}

\subsection{Ward identities in the quark sector}
In the context of conformal field theory (CFT), the decomposition of \eqref{dec1} reveals that all terms on the right-hand side - except for the transverse-traceless leading contribution -  arise as specific solutions to the transverse and trace Ward identities. These identities encode the effects of quantum anomalies, which originate from renormalization and are dictated by a single counterterm proportional to the squared field strength \( F^2 \), given by \eqref{cct}. The conservation Ward identities take the form 

\begin{align}
\label{tr}
p_{1\n_1}\braket{T^{\mu_1\nu_1}(p_1)\,J^{\m_2 a}(p_2)\,J^{\m_3 b}(p_3)}_q&=4\,\big[\d^{\m_1\m_2}p_{2\l}\braket{J^{\l a}(p_1+p_2)\,J^{\m_3 b}(p_3)}_q-p_2^{\m_1}\braket{J^{\m_2 a}(p_1+p_2)\,J^{\m_3 b}(p_3)}_q\big]\notag\\
&+4\,\big[\d^{\m_1\m_3}p_{3\l}\braket{J^{\l a}(p_1+p_3)\,J^{\m_2 b}(p_2)}_q-p_3^{\m_1}\braket{J^{\m_3 a }_q(p_1+p_3)\,J^{\m_2 b}(p_2)}_q\big],
\end{align}
while the gauge Ward identities are given by
\begin{align}
\label{x1}
p_{2\m_2}\braket{T^{\mu_1\nu_1}(p_1)\,J^{\m_2 a }(p_2)\,J^{\m_3 b}(p_3)}_q&=0\\
p_{3\m_3}\braket{T^{\mu_1\nu_1}(p_1)\,J^{\m_2 a }(p_2)\,J^{\m_3 b}(p_3)}_q&=0.
\end{align}
These two equations are homogeneous because the $TJ$ two-point function, which is expected on their right-hand sides, is trivially zero. The trace Ward identity, instead, includes the anomaly contribution 
\begin{equation}
\label{x2}
g_{\mu\nu} \langle T^{\mu\nu}(q)J^{\alpha a_1}(p_1)J^{\beta a_2}(p_2)\rangle_q=\beta u^{\alpha \beta  \, a_1a_2},
\end{equation}
with 
\bea
&&u^{\alpha\beta a_1 a_2}(p_1,p_2, a_1, a_2) = -\frac{1}{4}\int\,d^4x\,\int\,d^4y\ e^{ip_1\cdot x + i p_2\cdot y}\ 
\frac{\delta^2 \{F^a_{\mu\nu}F^{a\mu\nu}(0)\}} {\delta A_{\alpha}^{a_1}(x) A_{\beta}^{a_2}(y)}\vline_{A^a=0} \,
\label{locvar}\\
\eea
explicitly given by
\beq
u^{\a \b a b} (p_1,p_2)\equiv \delta^{a b}\left((p_1\cdot p_2) g^{\alpha\beta} - p_2^\alpha p_1^\beta\right).
\eeq
This tensor structure summarizes the conformal anomaly contribution, being the Fourier transform of the anomaly term at $O(g^2)$. Additional terms are present at the level of three and four external gluons, which are also part of the hierarchy of these equations, as discussed in \cite{Coriano:2024qbr}, but are not part of this analysis.\\
The 2-point function of two conserved vector currents $J_i$ $(i=2,3)$ shown above in \eqref{tr}, in any CFT in $d$ dimensions is given by 
\beqa
\langle J_2^{\alpha a}(p)J_3^{\beta b}(-p) \rangle =\delta^{a b}\delta_{\Delta_2\, \Delta_3}\left(c_{123} \Gamma_J \right)\pi^{\alpha\beta}(p) (p^2)^{\Delta_2-d/2},
\qquad \Gamma_J=\frac{\pi^{d/2}}{ 4^{\Delta_2 -d/2}}\frac{\Gamma(d/2-\Delta_2)}{\Gamma(\Delta_2)},
\eeqa
with $c_{123}$ an overall constant and $\Delta_2=d-1$ is the scaling dimension of the vector current. 
The divergence of this 2-point function is removed by the counterterm vertex $\Gamma_{\text{count}}$ introduced below in \eqref{cct1}, as we will explain. In our case $\D_2=\D_3=d-1$ and Eq. (\ref{tr}) then takes the form 
\begin{align}
\label{2point}
p_{1\mu_1}\braket{T^{\mu_1\nu_1}(p_1)\,J^{\m_2 a }(p_2)\,J^{\m_3 b}(p_3)}_q&=4\delta^{a b} c_{123} \Gamma_J
\left( \delta^{\nu_1\mu_2}\frac{p_{2\lambda}}{(p_3^2)^{d/2 -\Delta_2}} {\pi^{\lambda \mu_3}}(p_3) -
\frac{p_2^{\nu_1}}{(p_3^2)^{d/2 -\Delta_2}}\pi^{\mu_2\mu_3}(p_3) \right.\nn\\
& \left. + \delta^{\nu_1\mu_3}\frac{p_{3\lambda}}{(p_2^2)^{d/2-\Delta_2}}\pi^{\lambda \mu_2}(p_2) -\frac{p_3^{\nu_1}}{(p_2^2)^{d/2-\Delta_2}}\pi^{\mu_3\mu_2}(p_2)\right).
\end{align}\\

 \subsection{The tensorial sectors decomposition for quarks }
 \label{eight}
 
Using the projectors $\Pi$ and $\pi$ one can write the most general form of the transverse-traceless part as
	\begin{equation}
		{\braket{t^{\m \n}(p_1)\,j^{ \a \, a }(p_2)\,j^{\b \, b }(p_3)}}_q =\Pi^{\m \n}_{\m_1 \n_1}(p_1)\pi^{\a}_{\a_1 }(p_2)\pi^{\b}_{\b_1}(p_3)\,\,X^{ \, \m_1 \n_1 \a_1 \b_1 a b}_q,
	\end{equation}
	where $X^{a b \, \m_1 \n_1 \a_1 \b_1}_q$ is a general tensor of rank four built out of the metric and momenta. We can enumerate all possible tensor that can appear in $X^{a b \, \m_1 \n_1 \a_1 \b_1}$ preserving the symmetry of the 
	correlator 
	\begin{align}
		\langle t^{\mu \nu }(p_1)j^{a \, \a}(p_2)j^{b \, \b}(p_3)\rangle_q & =
		{\Pi_1}^{\mu \nu}_{\m_1 \n_1}{\pi_2}^{\a}_{\a_1}{\pi_3}^{\b}_{\b_1}
		\left( A_{1}^{(q) a b} \ p_2^{\m_1 }p_2^{\n_1}p_3^{\a_1}p_1^{\b_1} + 
		A_{2}^{(q) a b}\ \delta^{\a_1 \b_1} p_2^{\m_1}p_2^{\n_1} + 
		A_{3}^{(q) a b}\ \delta^{\m_1\a_1}p_2^{\n_1}p_1^{\b_1}\right. \notag\\
		& \left. + 
		A_{3}^{(q) a b}(p_2\leftrightarrow p_3)\delta^{\m_1\b_1}p_2^{\n_1}p_3^{\a_1}
		+ A_{4}^{(q) a b}\  \delta^{\m_1\b_1}\delta^{\a_1\n_1}\right)\label{DecompTJJ}
	\end{align}
	with the reconstruction taking the form 

	\begin{align}
& \la T_{\mu_1 \nu_1}(\bs{p}_1) J^{\mu_2 a_2}(\bs{p}_2) J^{\mu_3 a_3}(\bs{p}_3) \ra_q = \la t_{\mu_1 \nu_1}(\bs{p}_1) j^{\mu_2 a_2}(\bs{p}_2) j^{\mu_3 a_3}(\bs{p}_3) \ra_q \nn\\[0.5ex]
& \qquad +  2 \mathscr{T}_{\mu_1 \nu_1}^{\quad\,\,\alpha}(\bs{p}_1)\Big[\delta_{[\alpha}^{\mu_3}p_{3\beta]}  \la J^{\mu_2 a_2}(\bs{p}_2) J^{\beta a_3}(-\bs{p}_2) \ra_q
+\delta_{[\alpha}^{\mu_2}p_{2\beta]}  \la J^{\mu_3 a_3}(\bs{p}_3) J^{\beta a_2}(-\bs{p}_3) \ra_q\Big]
 \nn\\
& \qquad + \frac{1}{d - 1}\,\pi_{\mu_1 \nu_1}(\bs{p}_1) \mathcal{A}^{\mu_2 \mu_3 a_2 a_3}_q, \label{tjjdec}
\end{align}

where $\mathscr{T}_{\mu_1 \nu_1\alpha}$ is defined in \eqref{a:T} and $\langle JJ\rangle_q$ is the 2-point function of the gluon with a virtual quark
and
\beq
\mathcal{A}_q^{\alpha\beta ab} = -{1 \over 3} \, \,  \, \, \frac{g_s^2}{16\pi^2} \,  2n_f \delta^{ab}u^{\alpha \beta}(p_1,p_2),
\eeq
is the contribution of the anomaly to the expansion of the correlator. 

\subsection{Renormalization of the Ward Identities and the gluon sector}

The decomposition at one-loop of the gluon sector follows \eqref{dec1}, 
but its final expression is modified compared to 
\eqref{tjjdec}, which is affected by new trace contributions not present in the quark sector. The main difference come from the external identities constraining the correlator in the presence of a gauge-fixing sector in the QCD action. The Ward identities do not hold any longer and re replaced by Slavnov-Taylor identities, derived from the BRST symmetry. As shown in \cite{Coriano:2024qbr}, the relevant 
WIs and STIs that can be used in order to fix the expression 
The first Ward identity is similar to \eqref{tr}.
A second identity can be derived using the BRST symmetry of the gauge fixed action. For this purpose, it is sufficient to choose an appropriate Green's function with one stress energy tensor, a derivative of the gluon field and a ghost field,  $\langle T_{\mu\nu} \partial^{\alpha} A_{\alpha}^a \bar c^b \rangle$, and then use its BRST invariance to obtain
\cite{Coriano:2024qbr,Armillis:2010qk}
the identity
 \beq
\partial_{x_1}^{\alpha}\partial_{x_2}^{\beta}\langle T_{\mu\nu}(x) J_{\alpha}^a(x_1)J_{\beta}^b(x_2)\rangle = 0,
\eeq
which in momentum space becomes
\begin{equation}
p_2^{\alpha} p_3^{\beta} \langle T_{\mu\nu}(p_1) J_{\alpha}^a(p_2)J_{\beta}^b(p_3)\rangle = 0. \label{brstwi}
\end{equation}
Notice that this constraint is weaker compared to an ordinary Ward identity, since we require a vanishing constraint after contractions with both vector momenta. This allows additional tensor structures to be nonzero in the sector decomposition of the correlator, compared to the quark sector. This sets a distinction between the ordinary CFT analysis of this correlator, which imposes just ordinary non-Abelian Ward identities on the vector lines, compared to the free-field theory realization that we discuss in our work.  
The renormalization of the Ward/Slavnov-Taylor identities for the entire vertex is obtained by the counterterm  action
\beq
\label{cct}
\mathcal{S}_{\textrm{count}}=\frac{1}{(d-4)} \frac{g_s^2}{16 \pi^2}(\frac{5}{3}C_A-\frac{2}{3} n_f)
\int d^4 x g \sqrt{g} F^{a\, \mu\nu}F^{a}_{\mu\nu}.
\eeq
From this expression we extract the associated counterterm vertex $\Gamma_{\textrm{count}} (q,p_1,p_2)$ by functional differentiation with respect to the external fields and then Fourier transforming to momentum space
\beq
\label{cct1}
\Gamma^{\mu\nu\alpha\beta\, ab}_{\text{count}}(z,x,y) =  \frac{2}{\sqrt{-g}}\frac{\delta^3 S_{\textrm{count}}}{\delta{g_{\mu\nu}}\delta A^a_\alpha(x)\, \delta A^b_\beta(y)}\vline_{A = 0}.
\eeq
Using this counterterm all the hierarchical WIs (conservation, trace, etc. ) of $\Gamma +\Gamma_{\text{count}}$, with $\Gamma$
given in \eqref{res}, are renormalized at this order.\\
As shown in \cite{Coriano:2024qbr}, in free field theory the tensorial sector expansion is given by
\begin{equation}
\begin{aligned}
	& \langle T^{\mu \nu}(q)  J^{ a\alpha}(p_1) J^{ b\beta}(p_2)\rangle_g=\langle t^{\mu \nu}(q)  j^{ a\alpha}(p_1) j^{ b\beta}(p_2)\rangle_g +\langle t^{\mu \nu }(q)j_{loc}^{a  \a}(p_1)j^{b  \b}(p_2)\rangle_g +\langle t^{\mu \nu }(q)j^{a  \a}(p_1)j_{loc}^{b  \b}(p_2)\rangle_g\\
	& \qquad+2 \mathcal{I}^{\mu \nu \rho }(q)\left[\delta_{[\rho}^{\beta} p_{2 \sigma ]}\langle J^{ a\alpha}({p}_1) J^{ b\sigma}(-{p}_1)\rangle_g +\delta_{[\rho}^{\alpha} p_{1 \sigma]}\langle J^{ b\beta}({p}_2) J^{ a\sigma}(-p_2)\rangle\right]_g  
	+\frac{1}{d-1} \pi^{\mu \nu}(q) \left[\mathcal{A}^{\alpha \beta a b}_g+\mathcal{B}^{\alpha \beta a b}_g\right]
\end{aligned}
\label{res}
\end{equation}
with additional local terms of the form 
\beq
\langle t^{\mu \nu }(q)j^{\a \, a}(p_1)j_{loc}^{\b \, b}(p_2)\rangle_g. 
\label{adds}
\eeq
The contribution from the gluons to the anomaly at $O(\alpha_s)$ is given by
\begin{eqnarray}
 \mathcal{A}_g^{\alpha\beta ab} &=& {11 \over 3} \,  \frac{g_s^2}{16\pi^2} \, C_A  \delta^{ab}u^{\alpha\beta}(p_1,p_2).
\end{eqnarray}
As mentioned above, unlike the transverse traceless sector, the longitudinal sector of the decomposition retains additional contributions that do not vanish under Slavnov-Taylor identities, as they do in the quark sector or the general conformal solution discussed in 
\cite{Bzowski:2013sza}. These conditions, being homogeneous second-order differential constraints in coordinate space, are less restrictive than conventional Ward identities. Notice, however, that such terms are not part of the trace anomaly, since they are traceless by construction. Simultaneously, the trace sector undergoes further modifications due to additional terms linked to the gluon equations of motion, denoted here as \( \mathcal{B}^{\alpha \beta a b}_g \). These terms, absent in the on-shell decomposition, introduce corrections that alter the structure of the trace sector that define additional explicit breakings of the conformal symmetry, even for a vanishing quark mass. These are expected since, as already mentioned, the conformal symmetry of the one-loop expansion of the correlator is also violated by the gauge fixing condition.    
 The new longitudinal terms \eqref{adds} take the form 

 \begin{align}
	\langle t^{\mu \nu }(q)j^{\a \, a}(p_1)j_{loc}^{\b \, b}(p_2)\rangle_g & =
	{\Pi}^{\mu \nu}_{\m_1 \n_1}(q) \, \pi^\a _{\a_1} (p_1)\, {{p_2}_{\beta_1} p_2^\beta \over p_2^2}
	\left( 	B^{ab}_1 \, p_1^{\m_1} \, p_1^{\nu_1} \, p_2^{\a_1} \, p_2^{\b_1} + B_2^{ab} \, p_1^{\m_1} \, p_2^{\b_1} \, \delta^{\a_1 \n_1}  \right), 
\end{align}

which is orthogonal to the trace sector. Notice that these local (longitudinal) contributions vanish when the gluons are on-shell. 
\subsection{The complete correlator}
The complete expansion of the correlator is obtained by combining the two sectors. 
In the massless fermion limit, the perturbative $TJJ$ is given by the expression

\begin{equation}
\begin{aligned}
	\Gamma^{\mu\nu\alpha\beta a b}(q,p_1,p_2) & =\langle t^{\mu \nu}(q)  j^{ \a a}(p_1) j^{ \b b}(p_2)\rangle+\langle t^{\mu \nu }(q)j_{loc}^{\a  a}(p_1)j^{\b  b}(p_2)\rangle_g +\langle t^{\mu \nu }(q)j^{\a  a}(p_1)j_{loc}^{\b  b}(p_2)\rangle_g\\
	& \qquad+2 \mathcal{I}^{\mu \nu \rho }(q)\left[\delta_{[\rho}^{\beta} p_{2 \sigma ]}\langle J^{ a\alpha}({p}_1) J^{ b\sigma}(-{p}_1)\rangle+\delta_{[\rho}^{\alpha} p_{1 \sigma]}\langle J^{ b\beta}({p}_2) J^{ a\sigma}(-p_2)\rangle\right] \\
&	+\frac{1}{ 3 \, q^2} \hat\pi^{\mu \nu}(q) \left[\mathcal{A}^{\alpha \beta a b}+\mathcal{B}^{\alpha \beta a b}_g\right]
\end{aligned}
\label{res}
\end{equation}
with 
\beq
\mathcal{A}^{\alpha \beta a b}=\mathcal{A}^{\alpha \beta a b}_q + \mathcal{A}^{\alpha \beta a b}_g,
\eeq
and the trace anomaly given by the expression
\beq
\mathcal{A}^{\alpha\beta ab} = {1 \over 3} \, \,  \, \, \frac{g_s^2}{16\pi^2} \, (11C_A - 2 n_f) \delta^{ab}u^{\alpha \beta}(p_1,p_2).
\eeq
Eq. \eqref{res} shows that the structure of the effective vertex corresponding to the $TJJ$ 
correlator is modified compared to the ordinary CFT case, encountered in \eqref{tjjdec},
with modifications affecting both the longitudinal and the trace sectors. 
 The traceless sector in $\Gamma$ is given by
\begin{equation}
\begin{aligned}
 \langle T^{\mu \nu}(q)  J^{ a\alpha}(p_1) J^{ b\beta}(p_2)\rangle_{tls}=&
\langle t^{\mu \nu}(q)  j^{ a\alpha}(p_1) j^{ b\beta}(p_2)\rangle+\langle t^{\mu \nu }(q)j_{loc}^{a  \a}(p_1)j^{b  \b}(p_2)\rangle_g +\langle t^{\mu \nu }(q)j^{a  \a}(p_1)j_{loc}^{b  \b}(p_2)\rangle_g\\
	& \qquad+2 \mathcal{I}^{\mu \nu \rho }(q)\left[\delta_{[\rho}^{\beta} p_{2 \sigma ]}\langle J^{ a\alpha}({p}_1) J^{ b\sigma}(-{p}_1)\rangle+\delta_{[\rho}^{\alpha} p_{1 \sigma]}\langle J^{ b\beta}({p}_2) J^{ a\sigma}(-p_2)\rangle\right], 
\end{aligned}
\label{ali}
\end{equation}

and the trace part by

\beq
\label{anommX}
\langle T^{\mu \nu}(q)  J^{ a\alpha}(p_1) J^{ b\beta}(p_2)\rangle_{tr}=\frac{1}{ 3 \, q^2} \hat\pi^{\mu \nu}(q) \left[\mathcal{A}^{\alpha \beta a b}+\mathcal{B}^{\alpha \beta a b}_g\right],
\eeq
with
\beq
\mathcal{B}_g^{\alpha \beta\, a b }\equiv \delta^{a b} \mathcal{B}_g^{\alpha \beta }. 
\eeq
Therefore, in the conformal limit, from \eqref{anommX} we derive the trace anomaly condition

\beq
\label{vg}
g_{\mu\nu}\langle T^{\mu \nu}(q)  J^{ \a a}(p_1) J^{ \b b}(p_2)\rangle_{tr}= \left[\mathcal{A}^{\alpha \beta a b}+\mathcal{B}^{\alpha \beta a b}_g\right].
\eeq
The contraction with $g_{\mu\nu}$ has removed the anomaly poles from the trace sector, while the $\mathcal{B}_g$ in \eqref{vg} is uniquely introduced by the gauge-dependent contributions introduced by the gluon sector in the vertex. \\
These terms vanish for 
on-shell gluons and take the form
\beq
\mathcal{B}_g^{\alpha \beta}(q^2, p_1^2,p_2^2)=\left[ C_1 \, p_1^\alpha \, p_1 ^\beta + C_2 \, p_1^\alpha \, p_2^\beta + C_3  \, p_1^\beta \, p_2^\alpha + C_1(p_1\leftrightarrow p_2) \, p_2^\alpha \, p_2^\beta + C_4\, g^{\alpha \beta} \right].  
\label{bb}
\eeq
 They are generated by the explicit breaking of the classical conformal symmetry of the QCD action induced by the gauge fixing procedure.\\
The coefficient functions in \eqref{bb} $C_1,\ldots C_4$ are explicitly given in Appendix \ref{gg} and depend only on 
the virtualities of the external momenta. Notice that in \eqref{res} and \eqref{anommX}, the \({1}/{q^2} \) anomaly pole has been explicitly extracted from the longitudinal projector \( \pi^{\mu\nu} \) - the last term on the right-hand side of \eqref{res} - by defining
\beq
\hat\pi^{\mu\nu} \equiv q^2 g^{\mu\nu} - q^\mu q^\nu, \qquad  \pi^{\mu\nu}=\frac{1}{q^2}\hat\pi^{\mu\nu}. 
\eeq
In the tensorial decomposition of \eqref{res}, the pole emerges directly as a result of algebraic manipulations, at this stage without any particular physical interpretation. Its role becomes evident through an analysis of the sum rule associated with a specific tensor component of the trace sector and signals a dilaton exchange in the perturbative picture. \\ 
In the presence of fermion masses, the pole is smoothed into a momentum–dependent form factor, \({\Phi}_{\text{TJJ}}(q^{2},p_1^2,p_2^2,m^{2})\), which we analyze dispersively and which reproduces the \(1/q^{2}\) anomaly pole as \(m\to0\) and for 
on-shell gluons.
This form factor, here shown in the conformal limit, can then be defined in such a limit as 
\beq
\label{holds}
\Phi_{TJJ}\equiv \frac{\mathcal{A}^{\alpha\beta ab}}{q^2}. 
\eeq
The inclusion of the $1/q^2$ factor is essential for the saturation of the sum rule in the conformal limit, a point that we will address in the next sections.
In the conformal limit in which \eqref{holds} holds, this and other types of anomaly interactions manifest identical features. This occurs for chiral, conformal, and gravitational anomalies \cite{Giannotti:2008cv, Armillis:2010qk, Armillis:2009pq}, including the study of parity-violating trace anomalies (CFTs)~\cite{Coriano:2023hts}.\\
The term { anomaly pole}, clearly identified in the conformal limit, has been introduced to emphasize the subordinate nature of this interaction compared to that of a conventional particle pole, which corresponds to an asymptotic state. Its physical relevance becomes tied to the saturation of a sum rule that governs the absorptive part of the vertex. \\
This sum rule may be saturated by a pole, a branch cut, or both, depending on the external kinematical conditions and the presence or absence of mass corrections.
As we shall see, the identification of which terms contribute to the anomaly form factor in the general off-shell and massive case, as opposed to those that don't, is a subtle task that we are going to address. In fact, we will retain, in the definition of the anomaly form factor, a specific component of $\mathcal{B}_g$ which is proportional to $u^{\alpha\beta}$, the image of the anomaly in momentum space. This clearly implies that the anomaly form factor is gauge-dependent, but the contributions of these extra terms to the sum rule, as we are going to show, vanish. The sum rule is gauge-invariant and numerically related to $\beta(g)$.
As we move to the massive fermion case, the corresponding expression of \eqref{anommX} includes extra mass-dependent terms. These characterize the explicit breaking of conformal symmetry due to mass corrections coming from the quark sector, while the contribution of the gluon sector is already contained in \eqref{anommX} and remains unmodified.  
 
\section{Identifying the anomaly form factor for the massive correlator}
In order to move away from the conformal point and separate and identify the conformal anomaly form factor, we re-investigate the correlator 
by including a fermion mass $m$ in the perturbative expansion. \\
We proceed from the quark sector. The decomposition given in \eqref{res} remains valid; however, the trace sector takes a modified form, naturally splitting into two distinct contributions. We obtain the trace contribution
\begin{equation}
\label{sept}
    g_{\mu\nu} \braket{T^{\mu\nu}J^{\alpha a}J^{\beta b}}_q = \left(\phi_1^{\alpha\beta}(p_1,p_2,q,m) + \phi_2^{\alpha\beta}(p_1,p_2,q,m)\right) \delta^{ab},
    \end{equation}  
separated into two terms.
The first, denoted as \(\phi_1\), is directly projected onto the tensor structure \(u^{\alpha\beta}\), while the second one \(\phi_2\), as we are going to show, doesn't.  In $\phi_1$ we identify two contributions
\begin{equation}
   \phi_1^{\alpha\beta a b} = \left(-\frac{2}{3} n_f \frac{g_s^2}{16\pi^2} + \chi_0(p_1,p_2,q,m)\right) \delta^{ab} u^{\alpha\beta}(p_1,p_2),
\end{equation}  
containing the quark flavour number \(n_f\) contribution to the anomaly pole in the conformal limit, now supplemented by an additional mass-dependent term, $\chi_0$. The function \(\chi_0\) is given by
\begin{equation}
\chi_0(p_1,p_2,q,m) \equiv \frac{B_1 - B_4 }{2\, p_1\cdot p_2}, 
\end{equation}
where the scalar functions $B_1,\ldots B_4$ are given in the appendix. Explicitly
 
\begin{align}
\label{parm}
	\chi_0=&g_s^2m^4  H_1\, C_0 \left(p_1^2,p_2^2,q^2,m^2\right) +g_s^2m^2 \Bigl(H_2  \,C_0 \left(p_1^2,p_2^2,q^2,m^2\right)  \nn\\&+ H_3 \,\bar{B}_0(q^2,m^2)+ H_4 \,\bar{B}_0(p_1^2,m^2)+ H_5 \,\bar{B}_0(p_2^2,m^2) +H_6\,\Bigl).
\end{align}
The coefficients functions $H_1\ldots H_6$ ($H_i\equiv H_i(p_1^2,p_2^2,q^2))$
depend only on the virtualities of the external momenta and not on $m$.  $C_0 \left(p_1^2,p_2^2,q^2,m^2\right)$ denotes the massive scalar 3-point function and $\bar{B}_0(q^2,m^2)$ is the renormalized scalar massive 2-point function. Their expressions are given in Appendix \ref{Co}.\\
Concerning the tensor structure $\phi_2$ in \eqref{sept}, as just mentioned, it appears in the trace but does not generate terms proportional to the tensor structure $u^{\alpha\beta}$ and as such it is not included in the anomaly form factor. Indeed we obtain
\begin{equation}
\label{p2}
    \phi_2^{\alpha\beta}(p_1,p_2,q,m) = \chi_1(p_1,p_2,q,m) v^{\alpha\beta} + \left( B_2 p_1^{\alpha}p_1^{\beta} + B_2(p_1 \leftrightarrow p_2) p_2^{\alpha}p_2^{\beta} + B_3 p_1^{\alpha}p_2^{\beta} \right), 
 \end{equation}  
with
\begin{equation}
    v^{\alpha\beta}(p_1,p_2) = (p_1\cdot p_2) g^{\alpha\beta} + p_2^\alpha p_1^\beta
\end{equation} 
and
\begin{equation}
\chi_1(p_1,p_2,q,m) \equiv \frac{B_1 + B_4 }{2\, p_1\cdot p_2}.
\end{equation} 
$B_1\ldots B_4$ are given in Appendix \ref{appfunc}.
Explicitly 
\allowdisplaybreaks{
\begin{align}
	\chi_1=s_1\, s_2\,m^2\,g_s^2 n_f\,\delta ^{a b} \Biggl(&\frac{ \left(-s_2 (s+s_1)-(s-s_1)^2+2 s_2^2\right) {B}_0\left(s_2,m^2\right)}{6 \pi ^2 s (s-s_1-s_2) \left(s^2-2 s (s_1+s_2)+(s_1-s_2)^2\right)^2}
	\nn\\&+\frac{  \left(2 s^2-s (s_1+s_2)-(s_1-s_2)^2\right) {B}_0\left(s,m^2\right)}{6 \pi ^2 s (s-s_1-s_2) \left(s^2-2 s (s_1+s_2)+(s_1-s_2)^2\right)^2}
	\nn\\&+\frac{ \left(-s^2-s (s_1-2 s_2)+(s_1-s_2) (2 s_1+s_2)\right) {B}_0\left(s_1,m^2\right)}{6 \pi ^2 s (s-s_1-s_2) \left(s^2-2 s (s_1+s_2)+(s_1-s_2)^2\right)^2}
	\nn\\&+\frac{ \left(s^3-s^2 (s_1+s_2)-s \left(s_1^2-6 s_1 s_2+s_2^2\right)+(s_1-s_2)^2 (s_1+s_2)\right) {C}_0\left(s_1,s_2,s,m^2\right)}{12 \pi ^2 s (s-s_1-s_2) \left(s^2-2 s (s_1+s_2)+(s_1-s_2)^2\right)^2}
	\nn\\&+\frac{1}{6 \pi ^2 s (s-s_1-s_2) \left(s^2-2 s (s_1+s_2)+(s_1-s_2)^2\right)}	\nn\\&+\frac{m^2  {C}_0\left(s_1,s_2,s,m^2\right)}{3 \pi ^2 s (s-s_1-s_2) \left(s^2-2 s (s_1+s_2)+(s_1-s_2)^2\right)}\Biggl).
\end{align}
 }
In summary, the decomposition in \eqref{sept} is globally organized in terms of a projector onto the tensor structure $u^{\alpha\beta}$—which captures the genuine anomaly contribution—and a second structure, $v^{\alpha\beta}$, which is unrelated to the anomaly, identifying an explicit breaking. The scalar function $\chi_0$ accounts for the mass corrections to the anomaly in this sector.
The remaining terms in brackets in \eqref{p2} vanish if multiplied by transverse polarization vectors of the two gluons.
This separation is conceptually natural: in the quark sector, the anomaly condition admits a decomposition into the pure anomaly term and a mass-dependent contribution, the latter arising from the insertion of the scalar operator $\sim m \bar\psi \psi$ into the $JJ$ two-point function. \\
\subsection{Extension to the gluon sector}
This approach is naturally extended to the gluon sector. This sector is unaffected by the mass corrections. For this reason,  we organize the gluonic part of the trace $\mathcal{B}^{\alpha\beta}_g$ in a similar way, following the same template. Obviously, this second sector does not carry any mass term, but we require that the tensor decomposition in terms of $v^{\alpha\beta}$ and the remaining tensor structures quadratic in $p_i^{\alpha} p_j^{\beta}\, i,j=1,2$ be identical to that of the quark sector.  
Therefore the contribution to the trace of the correlator from the virtual gluon sector is organised in the form
\begin{equation}
	g_{\m\n}\braket{T^{\m\n}J^{\a a}J^{\b b}}_g=\left( \frac{g_s^2}{16\pi^2}\, \frac{11}{3}C_A  \right) u^{\a\b}\delta^{ab} +\mathcal{B}^{\alpha \beta}_g \delta^{ab},
\end{equation}
as extracted from \eqref{vg}. Also in this case, the contribution proportional to $u^{\alpha \beta}$ in $\mathcal{B}^{\alpha \beta}_g$ is identified by the decomposition
\begin{equation}
	 \mathcal{B}^{\alpha \beta a b}_g=\chi_g(p_1,p_2,q)u^{\a\b}\d^{ab}+\phi_g^{\alpha \beta a b}
\end{equation}
where only
\beq
\chi_g(p_1,p_2,q)\equiv \frac{C_4- (p_1\cdot p_2)C_3 }{2\, p_1\cdot p_2}
\label{chig}
\eeq
projects into the anomaly structure $u^{\a\b}$.
The remaining tensor component 
\begin{equation}
	\phi_g^{\alpha \beta a b}=\chi_g'(p_1,p_2,q)v^{\a\b}\d^{ab}+C_1 \, p_1^\alpha \, p_1 ^\beta \d^{ab}+ C_2 \, p_1^\alpha \, p_2^\beta  \d^{ab}+ C_1(p_1\leftrightarrow p_2) \, p_2^\alpha \, p_2^\beta\d^{ab}
\end{equation}
weither contains terms that do not project on $u^{\a\b}$  
\begin{equation}
	\chi_g'(p_1,p_2,q)\equiv \frac{C_4+ (p_1\cdot p_2)C_3 }{2\, p_1\cdot p_2}.
\end{equation}
or vanish by contracting with the polarization vectors of the gluons. The functions $C_i$, are given in Appendix \ref{gg}. \\
 The final decomposition of the trace of the $TJJ$ correlator then can be written as
\begin{align}
	\langle T^{\mu \nu}(q)  J^{ a\alpha}(p_1) J^{ b\beta}(p_2)\rangle_{tr}=\frac{1}{ 3 \, q^2}\hat\pi^{\mu \nu}(q) \biggl[& \, \,  \, \,  \Biggl(\frac{g_s^2}{16\pi^2} \, \bigg(\frac{11}{3}C_A - \frac{2}{3} n_f \bigg)+ \chi_0(p_1,p_2,q,m)+\chi_g(p_1,p_2,q)  \Biggl)\delta^{ab}u^{\alpha \beta}(p_1,p_2) \nn\\&+ \left(\phi_2^{\alpha \beta a b} +\phi_g^{\alpha \beta a b}\right) \biggl],
\end{align}
that allows to define the anomaly form factor as 
\beq
\Phi_{TJJ} (p_1^2,p_2^2,q^2,m^2)\equiv \frac{1}{3q^2}\mathcal{A}=\frac{1}{3 q^2}\Big(\frac{g_s^2}{48\pi^2}(11C_A - 2 n_f) + \chi_0(p_1,p_2,q,m)+\chi_g(p_1,p_2,q)\Big).
\eeq
Combining this result with \eqref{parm}, its expression can be rewritten in the form 
\begin{align}
\Phi_{TJJ}=\Bigl(&\frac{g_s^2}{144 \pi^2q^2}(11C_A - 2 n_f)  +\frac{1}{3q^2} g_s^2m^4 H_1\, C_0 \left(p_1^2,p_2^2,q^2,m^2\right) +\frac{1}{3q^2}g_s^2m^2 \Bigl(H_2\,C_0 \left(p_1^2,p_2^2,q^2,m^2\right)  \nn\\&+ H_3\, \bar{B}_0(q^2,m^2)+ H_4\, \bar{B}_0(p_1^2,m^2)+ H_5 \,\bar{B}_0(p_2^2,m^2) +H_6\,+\chi_g(p_1,p_2,q)\Bigl).
\label{tjj}
\end{align}
This defines the function that we are going to investigate in our spectral analysis. 
Notice that the the first $1/q^2$ contribution in \eqref{tjj} has a residue given by the QCD $\beta$ function up to a sign $\sim 1/3 (11C_A - 2 n_f)$. With the definition above, the trace sector can be expressed in terms of $\Phi_{TJJ}$ as 
\begin{align}
\label{last}
	\langle T^{\mu \nu}(q)  J^{ a\alpha}(p_1) J^{ b\beta}(p_2)\rangle_{tr}=&\hat\pi^{\mu \nu}(q)\delta^{ab}\Bigl[ \Phi_{TJJ}(q)\, u^{\alpha\beta}(p_1,p_2) \nn \\
	& +\frac{1}{3}\left( \chi_1(p_1,p_2,q,m) 
	+ \chi'_g(p_1,p_2) \right)  v^{\alpha\beta}(p_1,p_2) \nn\\
	& +\frac{1}{3} \left((B_2 + C_1) p_1^{\alpha}p_1^{\beta} + \left[ B_2(p_1 \leftrightarrow p_2) + C_1(p_1 \leftrightarrow p_2)\right] p_2^{\alpha}p_2^{\beta} + ( B_3 + C_2) p_1^{\alpha}p_2^{\beta}\right) \Bigl],
\end{align}
where we have separated the contribution projecting onto the anomaly term $F^2$ from the remaining terms
that are part of the trace but not of the anomaly. The remaining terms, not included in the definition of the form factor, are those proportional to the tensor structure
$v^{\alpha b}$ and—in the last bracket of \eqref{last}—terms that vanish once contracted with the external polarizations of the gluons. We will show, through a detailed analysis of its spectral density, that the spectral density of this form factor is characterized by a sum rule valid for generic virtualities of the external gluons.\\
Notice that $\chi_g$ in \eqref{chig}, originating from the gluon sector, is a gauge-dependent contribution that must necessarily be included in the definition of the anomaly form factor when performing a dispersive analysis for off-shell gluons.

\section{Derivation of the sum rule}
Having identified the anomaly form factor, we can proceed with the analysis of the spectral function, proving the sum rule satified by its spectral density.\\  
The analyticity properties of the function \(\Phi_0\equiv \Phi_{TJJ}\) in the complex \(q^2\)-plane  \(\Phi_0\) satisfies relation of the form
\beq
\Phi_0(q^2, s_1, s_2, m^2) = \frac{1}{2\pi i} \oint_C \frac{\Phi_0(s, s_1, s_2, m^2)}{s - q^2}\, ds,
\eeq
with $s_1$ and $s_2$ real, where \(C\) is a closed contour encircling the singularities of \(\Phi_0\). The integration can be deformed to run over the $s=q^2 \geq 0$ on the real axis. 
As we will demonstrate, \(\Phi_0\) possesses both poles and a branch cut on the real axis, with the cut beginning at the physical threshold \(s = 4m^2\). Correspondingly, the associated spectral density will reflect this analytic structure, featuring poles from rational terms and a discontinuity across the cut for \(s > 4m^2\).\\
We begin by characterizing the discontinuity of the function \(\Phi_0\) across the branch cut for time-like momenta \(q^2 > 0\), defined as:
\beq
\label{disc}
\mathrm{Disc}\,\Phi_0(q^2) = \Phi_0(q^2 + i\epsilon, p_1^2, p_2^2, m^2) - \Phi_0(q^2 - i\epsilon, p_1^2, p_2^2, m^2),
\eeq
which encodes the non-analytic behavior of \(\Phi_0\) in the complex \(q^2\)-plane. 
\\
The spectral density of $\Phi_0$, denoted by \(\Delta \Phi_0\), is then introduced via the relation
\beq
\label{delta}
\Delta\Phi_0(s, s_1, s_2, m^2) \equiv \mathrm{Im}\,\Phi_0(s, s_1, s_2, m^2) = \frac{1}{2i} \mathrm{disc}\,\Phi_0(s).
\eeq
The contribution from the boundary at infinity vanishes and the dispersion relation simplifies to
\beq
\Phi_0(q^2, p_1^2, p_2^2, m^2) = \frac{1}{\pi} \int_0^\infty \frac{\Delta\Phi_0(s, p_1^2, p_2^2, m^2)}{s - q^2} \, ds.
\eeq
Coming to the spectral function of this form factor, we separate the discontinuity into the pole contributions plus the continuum, in the form 
\begin{equation}
	\text{disc}(\Phi_{TJJ})=\text{disc}(\Phi_{TJJ})_{cont}+\text{disc}(\Phi_{TJJ})_{pole},
\end{equation}
with the discontinuity of \eqref{tjj} over the branch cut given by 
\begin{align}
	\label{cont}
	\text{disc}(\Phi_{TJJ})_{cont}=&\frac{n_F\,g_s^2\,  m^2 K_1(s,s_1,s_2)\, \text{disc}\, {B}_0\left(s,m^2\right)}{12\pi ^2  s(s-s_1-s_2) \lambda^2}+\frac{n_F\,g_s^2\,  m^2  K_2(s,s_1,s_2) \,\text{disc}\,{C}_0\left(s,s_1,s_2,m^2\right)}{24\pi ^2  s(s-s_1-s_2) \lambda^2}\nn\\&\frac{C_A g_s^2 (s_1+s_2)\,\text{disc}\,  {B}_0(s)}{48 \pi ^2 \l}+\frac{C_A g_s^2\, K_3\, \text{disc}\, {C}_0\left(s,s_1,s_2\right)}{96 \pi ^2 s (s-s_1-s_2) \l},
\end{align}
with
\begin{align}
	K_1=-s_2 (s-s_1) \left(s^2+3 s s_1-2 s_1^2\right)-s_2^3 (3 s+2 s_1)+s_2^2 (s+s_1) (3 s+2 s_1)-s_1 (s-s_1)^3+s_2^4
\end{align}
\begin{align}
	K_2=&-4 m^2 \left(s^2-2 s (s_1+s_2)+(s_1-s_2)^2\right) \left(s^2-2 s (s_1+s_2)+s_1^2+s_2^2\right)\nn\\&+s_2^4 (5 s+s_1)-2 s s_2^3 (5 s+3 s_1)+2 s s_2^2 (s+s_1) (5 s-3 s_1)\nn\\&-s_2 (5 s-s_1) (s+s_1) (s-s_1)^2+(s-s_1)^5-s_2^5
\end{align}
\begin{align}
	K_3=&\,\,s_2^2 (-9 s^2+3 s s_1-10 s_1^2)+s_2 (3 s^3-10 s^2 s_1+3 s s_1^2+8 s_1^3)\nn\\&+s_2^3 (9 s+8 s_1)+3 s_1 (s-s_1)^3-3 s_2^4\, .
\end{align}
Notice that the contributions to the cuts or poles in the spectral density originating from the quark and gluon sectors can be traced separately by examining the coefficients of $n_f$ and $C_A$ in \eqref{cont}, respectively.\\
The general discontinuity of the scalar 3-point function $C_0$, given in Appendix \ref{Co}, is given by
\begin{equation}
	\label{d2}
	\text{disc}\,{C_0}(s,s_1,s_2,m^2)=-\frac{2i  \pi  \log \left(\frac{m^2-s (w-\bar z) (\bar w-z)}{m^2-s (w-z) (\bar w-\bar z)}\right)}{s (z-\bar z)}\,\,\, \theta(s-4m^2),
\end{equation}
where	
	
	\beq
		z=\frac{1}{2 s } \left(\sqrt{\lambda} + s_1 - s_2 +s\right) \qquad \bar z=z=\frac{1}{2 s } \left(- \sqrt{\lambda} + s_1 - s_2 +s\right)
\eeq
with
	\beq
		w=\frac{1}{2} \left(1+\sqrt{1-\frac{4 m^2}{s}}\right)	\qquad \bar w=\frac{1}{2} \left(1-\sqrt{1-\frac{4 m^2}{s}}\right),
\eeq	
and 
\beq
\text{disc}\, {C}_0\left(s,s_1,s_2\right)
= - \frac{2i\pi}{\sqrt{\lambda(s, s_1, s_2)}}
\log \left(
\frac{(2s - s_1 + s_2 + \sqrt{\lambda})(s_1 - s_2 + s + \sqrt{\lambda})}
     {(2s - s_1 + s_2 - \sqrt{\lambda})(s_1 - s_2 + s - \sqrt{\lambda})}
\right),
\eeq

while 
\begin{equation}
\label{bos}
	\textrm{disc}\,{B_0}(s,m^2)=2\p i \sqrt{1-\frac{4m^2}{s} }\,\,\,
	\theta(s-4m^2).
\end{equation}
The pole contributions split at separate locations $(s=0, s=s_1+s_2, s=s_\pm)$
\begin{align}
	\text{disc}(\Phi_{TJJ})_{pole}=\text{disc}(\Phi_{TJJ})_{0}+\text{disc}(\Phi_{TJJ})_{s_1+s_2}+\text{disc}(\Phi_{TJJ})_{s_-}+\text{disc}(\Phi_{TJJ})_{s_+},
\end{align}
where we have indicated the points corresponding to the zeroes of the $\lambda$ function as

 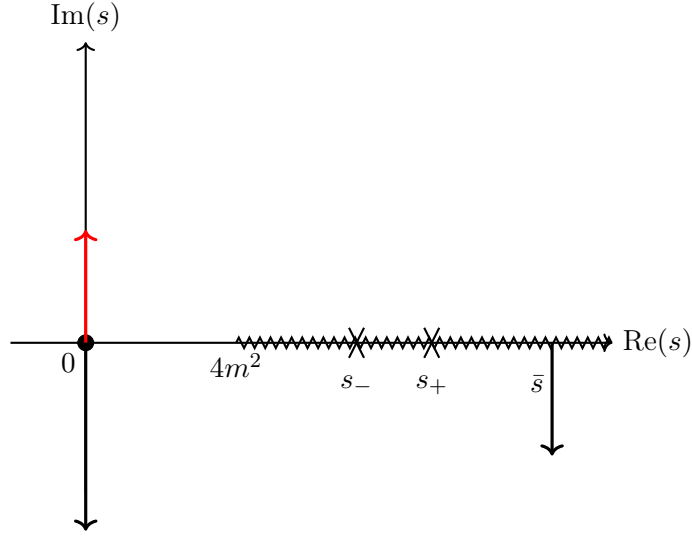
\begin{figure}
	\centering
	\begin{tikzpicture}
	
	\draw[thick, ->] (-1,0) -- (7,0) node[right] {$\text{Re}(s)$};
	\draw[thick, ->] (0,-2) -- (0,4) node[above] {$\text{Im}(s)$};
	
	\filldraw[black] (0,0) circle (3pt) node[below left] {$0$};
	
	\draw[thick, decorate, decoration={zigzag, amplitude=2pt, segment length=4pt}] (2,0) -- (7.0,0);
	\node at (2, -0.3) {$4m^2$};
	
	\draw[very thick, ->, red] (0,0) -- (0,1.5);     
	\draw[very thick, ->, black] (0,0) -- (0,-2.5); 
	
	\draw[thick] (3.5,0.2) -- (3.7,-0.2); 
	\draw[thick] (3.7,0.2) -- (3.5,-0.2); 
	\node[below] at (3.6,-0.3) {$s_-$};   
	
	\draw[thick] (4.5,0.2) -- (4.7,-0.2); 
	\draw[thick] (4.7,0.2) -- (4.5,-0.2); 
		\node[below] at (4.6,-0.3) {$s_+$};   
	\draw[very thick, ->, black] (6.2,0.0) -- (6.2,-1.5); 
         \node[below] at (6.0,-0.3) {$\bar s$}; 
        \end{tikzpicture}
	\caption{The spectral density of the off-shell conformal anomaly form factor in the complex \( s\equiv q^2 \)-plane. Shown are the two spikes at \(s=0\), the anomalous thresholds at $s_{\pm}=(\sqrt{s_1}\pm \sqrt{s_2})^2$ and the pole at $s=\bar{s}=s_1+s_2$. In this case there is a cancelation in the sum rule between the integral of the density over the cut, the black spike at $s=0$, of variable strength, and the black spike at $s=\bar{s}$.    } 
\label{figgX}
\end{figure}
\beq
s_\pm=(\sqrt{s_1} \pm \sqrt{s_2})^2.
\eeq
One extra pole is present at $s= s_1 + s_2$, beside the pole at $s=0$.
The poles at $s=0$ are split into two spikes, one of constant and the other of variable strength, represented in 
Fig. 3 
\begin{align}
	\label{Ym}
	\text{disc}(\Phi_{TJJ})_{0}=2\p i&\delta(s){ \, \,   \frac{g_s^2}{144\pi^2} \, (11C_A - 2 n_f)}-2\p i\delta(s)\Y(s_1,s_2,s, m^2)
\end{align}
where 
\allowdisplaybreaks{
\begin{align}
	\Psi(s_1,s_2,s=0, m^2)=&g_s^2\,\,C_A\,  \left(-\frac{s_1 s_2 {B}_0^R(s_1,0)}{24 \pi ^2 \left(s_1^2-s_2^2\right)}+\frac{s_1 s_2 {B}_0^R(s_2,0)}{24 \pi ^2 \left(s_1^2-s_2^2\right)}+\frac{\left(3 s_1^2-2 s_1 s_2+3 s_2^2\right) {C}_0(s_1,s_2,0)}{96 \pi ^2 (s_1+s_2)}+\frac{1}{48 \pi ^2}\right)\nn\\&
	+g_s^2\,\,n_F \, \Biggl(m^2 \biggl(-\frac{\left(s_1^2+s_2^2\right) {B}_0^R\left(0,m^2\right)}{12 \pi ^2 (s_1-s_2)^2 (s_1+s_2)}+\frac{s_1 \left(s_1^2+2 s_1 s_2+3 s_2^2\right) \text{B}_0^R\left(s_1,m^2\right)}{12 \pi ^2 (s_1-s_2)^3 (s_1+s_2)}\nn\\&
	+\frac{s_2 \left(3 s_1^2+2 s_1 s_2+s_2^2\right) {B}_0^R\left(s_2,m^2\right)}{12 \pi ^2 (s_2-s_1)^3 (s_1+s_2)}+\frac{\left(s_1^2+s_2^2\right) {C}_0\left(s_1,s_2,0,m^2\right)}{24 \pi ^2 (s_1-s_2)^2}+\frac{s_1^2+s_2^2}{12 \pi ^2 (s_1-s_2)^2 (s_1+s_2)}\biggl)\nn\\&
	+\frac{m^4 \left(s_1^2+s_2^2\right) {C}_0\left(s_1,s_2,0,m^2\right)}{6 \pi ^2 (s_1-s_2)^2 (s_1+s_2)}\Biggl).
	\label{psi}
\end{align}

where  $C_0(s_1,s_2,0)$ (in three variables) is the massless $C_0(s_1,s_2,s)$ integral evaluated at $s=0$, and 
${C}_0(s_1,s_2,0,m^2)$ is the massive $C_0(s_1,s_2, s,m^2)$ integral evaluated at the same point. We have denoted with 
We have outlined the various contributions in \figref{figgX}. The "black spike" at \( s = 0 \), which maintains a constant strength, corresponds to the first term on the right-hand side of \eqref{Ym}. The second contribution, which is dependent on the external invariants, is represented by the "red spike" in the same figure. \\
For \( s > 4m^2 \), across the cut, we encounter two apparent poles. As we will demonstrate, these poles do not contribute to the discontinuity. Additionally, there is a pole at \( s = s_1 + s_2 \), which can appear either above or below the cut (i.e., for \( s > 4m^2 \) or \( s < 4m^2 \)), depending on the specific values of \( s_1 \) and \( s_2 \).
Finally, we observe that in the limit \( s \to 0 \)
\begin{align}
		\lim_{s\to 0} \Y(s,s_1,s_2)=&g_s^2\,\,C_A\,  \left(-\frac{s_1 s_2 {B}_0^R(s_1,0)}{24 \pi ^2 \left(s_1^2-s_2^2\right)}+\frac{s_1 s_2{B}_0^R(s_2,0)}{24 \pi ^2 \left(s_1^2-s_2^2\right)}+\frac{\left(3 s_1^2-2 s_1 s_2+3 s_2^2\right){C}_0(s_1,s_2,0)}{96 \pi ^2 (s_1+s_2)}+\frac{1}{48 \pi ^2}\right)\nn\\&
		+g_s^2\,\,n_F \, \Biggl(m^2 \biggl(-\frac{\left(s_1^2+s_2^2\right){B}_0^R\left(0,m^2\right)}{12 \pi ^2 (s_1-s_2)^2 (s_1+s_2)}+\frac{s_1 \left(s_1^2+2 s_1 s_2+3 s_2^2\right){B}_0^R\left(s_1,m^2\right)}{12 \pi ^2 (s_1-s_2)^3 (s_1+s_2)}\nn\\&
		+\frac{s_2 \left(3 s_1^2+2 s_1 s_2+s_2^2\right) {B}_0^R\left(s_2,m^2\right)}{12 \pi ^2 (s_2-s_1)^3 (s_1+s_2)}+\frac{\left(s_1^2+s_2^2\right) {C}_0\left(s_1,s_2,0,m^2\right)}{24 \pi ^2 (s_1-s_2)^2}+\frac{s_1^2+s_2^2}{12 \pi ^2 (s_1-s_2)^2 (s_1+s_2)}\biggl)\nn\\&
		+\frac{m^4 \left(s_1^2+s_2^2\right) {C}_0\left(s_1,s_2,0,m^2\right)}{6 \pi ^2 (s_1-s_2)^2 (s_1+s_2)}\Biggl)  < \infty.
\end{align}
Notice that \(\Y\) remains finite across the entire light-cone for fixed $s_1$ and $s_2$.

\subsection{Cancellation of the anomalous thresholds and the sum rule}
\label{trash}
The expression for the spectral density of $\Phi_{TJJ}$ can be separated into distinct components, representing contributions from regular kinematical thresholds as well as potential anomalous thresholds. Therefore we split the form factor as
\begin{equation}
	\Phi_{{TJJ}} = \Phi_{{TJJ}}^{\text{reg}} + \Phi_{{TJJ}}^{(\l)},
\end{equation}
where $\Phi_{{TJJ}}^{\text{reg}}$ corresponds to regular kinematical thresholds and $\Phi_{{TJJ}}^{\l}$ accounts for possible anomalous thresholds in $\Phi_{{TJJ}}$. The regular part, $\Phi_{{TJJ}}^{\text{reg}}$, includes contributions from standard kinematical thresholds located at
\begin{equation}
	q^2 = 0, \quad q^2 = s_1 + s_2, \quad \text{and} \quad q^2 = 4m^2,
\end{equation}
which manifest in the form of simple poles and cuts. These thresholds arise from well-understood physical processes. In contrast, the anomalous part, $\Phi_{{TJJ}}^{\text{anom}}$, emerges under special conditions when the kinematical factor $\lambda = 0$ which could be present in this computation. Given that
\begin{equation}
	\lambda = (s - s_-)(s - s_+),
\end{equation}
the dispersive representation of  $ \Phi_{\text{TJJ}}$ may allow terms of the form
\begin{align}
	\textrm{disc} \,\Phi_{{TJJ}} = &c_0 \delta(s) + c_1(s,s_1, s_2) \theta(s - 4m^2) H(s) + c_2(s,s_1, s_2) \delta(s - s_1 - s_2)\nn\\&
	+ c_3(s, s_1, s_2) \delta'(s - s_-) + c_4 \delta'(s - s_+) + c_5 \delta(s - s_-) + c_6 \delta(s - s_+),
\end{align}
where the derivative of the $\delta$-function corresponds to higher order poles at $\lambda = 0$.
Indeed, the general structure of $\Phi_{\text{TJJ}}$ is
\begin{align}
	\Phi_{{TJJ}} =&  \frac{c_0(s,s_1,s_2,m^2)}{s} + \frac{c_1(s,s_1,s_2,m^2)}{s} + \text{(continuum terms)} + \frac{c_2(s,s_1,s_2,m^2)}{s - \bar{s}} + \frac{c_3(s,s_1,s_2,m^2)}{(s - s_-)^2} \nn\\&+ \frac{c_4(s,s_1,s_2,m^2)}{(s - s_+)^2} + \frac{c_5(s,s_1,s_2,m^2)}{s - s_-} + \frac{c_6(s,s_1,s_2,m^2)}{s - s_+},
\end{align}
where $\bar{s} = s_1 + s_2$ and, $c_3$ and $c_4$ define possible residues at the double poles of $\lambda^2$. These could, in principle, introduce subtractions that can be analyzed as follows. \\
For example, at $s = s_-$, we have
\begin{equation}
	\Phi_{{TJJ}}^{(\l)}(q^2 = s_-) = \frac{1}{2\pi i} \int_0^\infty \frac{c_3(s, s_1, s_2, m^2) \delta'(s - s_-)}{(s - q^2)} \, ds
	+ \frac{1}{2\pi i} \int_0^\infty \frac{c_5(s, s_1, s_2, m^2) \delta(s - s_-)}{(s - q^2)} \, ds.
\end{equation}
This simplifies to:
\begin{equation}
	{2\pi i}\,\Phi_{{TJJ}}^{(\l)}(q^2 = s_-)= -\frac{d}{ds} \left( \frac{c_3(s, s_1, s_2, m^2)}{(s - q^2)} \right) \bigg|_{s = s_-} - \frac{c_5(s_-, s_1, s_2, m^2)}{s_- - q^2}.
\end{equation}
A similar analysis gives 
\begin{equation}
	{2\pi i}\,\Phi_{{TJJ}}^{(\lambda)}(q^2 = s_+) =  -\frac{d}{ds} \left( \frac{c_4(s, s_1, s_2, m^2)}{(s - q^2)} \right) \bigg|_{s = s_+} - \frac{c_6(s_+, s_1, s_2, m^2)}{s_+ - q^2}.
\end{equation}
For the evaluation of $\Phi_{{TJJ}}^{(\lambda)}(q^2 = s_\pm)$, we need to compute the explicit expression of the scalar integral $C_0$ at these kinematical points. We use the expressions of the scalar integral $C_0$ at these points 
($C_0(s_-,s_1,s_2,m^2)$ and $C_0( s_+,s_1,s_2,m^2)$). Their expressions can be found in appendix \ref{Co}.

Using these relations one can show by a long computation that 

\begin{equation}
	\Phi_{{TJJ}}^{(\l)}(q^2 = s_\pm) \equiv 0,
\end{equation}
and hence 
\begin{equation}
	\Phi_{{TJJ}}^{(\l)}=0.
\end{equation}
Therefore, the only pole to consider at finite $s$ is the one located at $s= s_1+s_2$.
This contributes to the sum rule with a localized density that is given by 
{\small\begin{align}
		\text{disc}(\Phi_{TJJ})_{s_1+s_2}=&g_s^2\,\,C_A  \left(\frac{(s_2-s_1) \text{B}_0^R(s_1)}{96 \pi ^2 (s_1+s_2)}+\frac{(s_1-s_2) {B}_0^R(s_2)}{96 \pi ^2 (s_1+s_2)}-\frac{\left(s_1^2-6 s_1 s_2+s_2^2\right) {C}_0(s_1,s_2,s_1+s_2)}{96 \pi ^2 (s_1+s_2)}-\frac{1}{48 \pi ^2}\right)\nn\\&
		+g_s^2 \,\, n_F \Biggl(m^2 \biggl(-\frac{{B}_0^R\left(s_1,m^2\right)}{48 \pi ^2 (s_1+s_2)}+\frac{{B}_0^R\left(s_1+s_2,m^2\right)}{24 \pi ^2 (s_1+s_2)}-\frac{{B}_0^R\left(s_2,m^2\right)}{48 \pi ^2 (s_1+s_2)}+\frac{\text{C}_0\left(s_1,s_2,s_1+s_2,m^2\right)}{48 \pi ^2}\nn\\&-\frac{1}{24 \pi ^2 (s_1+s_2)}\biggl)-\frac{m^4 {C}_0\left(s_1,s_2,s_1+s_2,m^2\right)}{12 \pi ^2 (s_1+s_2)}\Biggl).
		\label{bo}
\end{align}}
Notice that the UV divergences in the $B_0$'s in the expression above cancel automatically, in the counting of the $1/\epsilon_{UV}$ poles in each of these self-energy contributions. Therefore we have replaced in \eqref{bo} each $B_0$ by $B_0^R$ with no change. No IR singularities are encountered unless we consider special values where any of the external invariants vanishes. 
\\
The mutual cancellation of some of the discrete contributions with the continuum, separately for the $n_f$ and $C_A$ terms, after integration over $s$, can be shown by a direct computation. One can verify the nontrivial relation 
\beq
\int_{4 m^2}^\infty d s\, \text{disc}(\Phi_{TJJ})_{cont}- 2 \pi i\psi(s_1,s_2,s=0,m^2) + \text{disc}(\Phi_{TJJ})_{s_1+s_2}=0,
\eeq
where we have used \eqref{Ym}, the expression of $\psi(s_1,s_2,s=0,m^2)$ defined in \eqref{psi}, \eqref{bo} and the continuum contribution given in \eqref{cont}. Using \eqref{Ym} one is clearly left only with the anomaly pole contribution at $s=0$ (the first term of \eqref{Ym}), giving 

\beq
\int _{0}^{\infty} ds \,\Delta \Phi_{TJJ}(s_1,s_2,s,m^2)=-\frac{1}{3}\frac{\beta(g)}{g},
\eeq
which is mass-independent and gauge invariant. This equation is the sum rule satisfied by the absorptive part of the conformal anomaly form factor in QCD at $O(\alpha_s)$.

\section{Sum rules of anomaly form factors in the on-shell limit }
In the on-shell case, the sum rules take a simpler form, and both the integral of the spectral density and the relation between the continuum and the anomaly pole exhibit a clearer pattern. While the chiral and conformal anomaly sum rules share structural similarities, they are not identical. To highlight the distinction between the two, we first summarize the main results of the chiral case before turning to the conformal case \cite{Coriano:2025ceu}.

\subsection{The chiral case: review}
We consider the $AVV$ correlator (axial-vector/vector/vector) and illustrate how the branch cut of the spectral density of its anomaly form factor evolves into a pole as the fermion mass approaches zero.
To explore deviations from conformal invariance in the structure of this form factor one considers the case of massive fermions and computes the correlator using standard perturbative technique. The $AVV$ correlator is constructed with two vector currents $J$ and one axial-vector current $J_A$, all Abelian, and can be parameterized in the form 
\begin{align}
\label{refm}
	\braket{J^{\mu_1}(p_1)J^{\mu_2}(p_2) J^{\mu_3}_A (p_3)} &=	\braket{J^{\mu_1 }(p_1) J^{\mu_2 }(p_2)\, j_{A \text { loc }}^{\mu_3}(p_3)}
 	+ \pi^{\mu_1}_{\alpha_1}(p_1)\pi^{\mu_2}_{\alpha_2} (p_2) \pi^{\mu_3}_{\alpha_3}(p_3)
	\, \Delta_T^{\alpha_1 \alpha_2 \alpha_3 }(p_1^2,p_2^2,p_3^2,m^2) 
	 \nonumber \\
\end{align}
with the longitudinal sector given by 
\begin{equation}
\label{nonloc}
		\braket{J^{\mu_1 }(p_1) J^{\mu_2 }(p_2)\, j_{A \text { loc }}^{\mu_3}(p_3)}\equiv \bar \Phi_0(p_1^2,p_2^2,p_3^2,m^2) \varepsilon^{p_1p_2\mu_1\mu_2}\,p_3^{\mu_3}.
	\end{equation}
We have used a decomposition based on the inclusion of projectors as in \eqref{loct}.	
  The expression of the transverse part is denoted by $\Delta_T$ and will not be relevant for our current analysis. A direct computation of the longitudinal part gives
\beq
\label{fst}
\bar \Phi_0(p_1^2,p_2^2,p_3^2,m^2)=\frac{ m^2}{ \pi ^2}\,\frac{1}{q^2}C_0\left(q^2,p_1^2 \,,p_2^2\,, m^2\right)+\frac{1 }{2 \pi ^2}\,\frac{1}{q^2},
\eeq	
whose discontinuity defined as in \eqref{disc}, for $s_1\equiv p_1^2=0,\, s_2=p_2^2=0$ - defining the on-shell case - takes the form 
\beqa
 \label{zero}
 \textrm{disc}\,\bar \Phi_0 (q^2, m^2) 
 &=&  \frac{m^2}{ \pi^2} \textrm{disc}\left( \frac{C_0(q^2,m^2)}{q^2}\right) +\frac{1}{2 \pi^2} \textrm{disc}\frac{1}{q^2},
 \eeqa
 and can be re-expressed as
 \begin{equation}
\mathrm{disc} \, \bar \Phi_0(q^2,m^2) =-\frac{1}{2\pi^2} \left(2 \pi i \, \delta(q^2)
+ 4 i \pi m^2 \, C_0(q^2 \to 0,m^2) \delta(q^2) - \frac{2 m^2}{q^2} \mathrm{disc} \, C_0(q^2, m^2)\right).
\label{second}
\end{equation}
There are three contributions to this function:
 two spikes at $q^2\equiv s=0$, represented by two uparrows, one of constant and the other of varaible strength, and a cut covering the $q^2>4 m^2$ 
 region.
 Introducing the limit 
 \begin{equation}
C_0(q^2 \to 0,m^2) = -\frac{1}{2 m^2}
\end{equation}
one notices the cancellation of the two localized contributions at $s = 0$. 
The anomaly form factor does not exhibit any pole at $p_3^2=0$, but has only a branch cut  for $q^2\equiv p_3^2 > 4\, m^2$. A similar result can be obtained form the explicit expression of $\Phi_0$, that can be derived in the form 
\begin{equation}
\bar \Phi_0= \frac{1}{2 \pi^2 q^2} \left( 1 + \frac{m^2}{q^2} \log^2 \left( \frac{\sqrt{\tau(q^2, m^2)} + 1}{\sqrt{\tau(q^2, m^2)} - 1} \right) \right), \qquad q^2 > 0,
\label{a6}
\end{equation}
with $\tau(q^2, m^2) = 1 - \frac{4m^2}{q^2}$ with
\begin{align}
\mathrm{disc} \, {C}_0(q^2, m^2) &= \frac{1}{i \pi^2} 
\int d^4 l \frac{2 \pi i \delta_+(l^2 - m^2) 2 \pi i \delta_+((l - q)^2 - m^2)}{(l - p_1)^2 - m^2 + i \epsilon} \notag \\
&= \frac{2 \pi}{i q^2} \log \left( \frac{1 + \sqrt{\tau(q^2, m^2)}}{1 - \sqrt{\tau(q^2, m^2)}} \right) \theta(q^2 - 4 m^2)
\label{cut}
\end{align}
that gives
\beq
\label{twist}
\textrm{disc} \, \bar \Phi_0(q^2, m^2) =-2 i  \frac{m^2}{\p (q^2)^2}\log \frac{1 + \sqrt{\tau(q^2,m^2)}}{1 - \sqrt{\tau(q^2,m^2)} }\theta(q^2- 4 m^2).
 \eeq 

Then, the final spectral density takes the form
\begin{equation}
\Delta {\bar \Phi_0}(q^2,m^2) = \frac{ m^2}{\pi s^2} \log \left( \frac{1 + \sqrt{\tau(s, m^2)}}{1 - \sqrt{\tau(s, m^2)}} \right) \theta(s - 4 m^2). 
\label{exp}
\end{equation}
and is only given by the contribution coming from the cut. 
A direct integration of this function gives 
\begin{equation}
\label{sumrr}
\frac{1}{\pi}\int_{4 m^2}^\infty ds\, \Delta{\bar \Phi_0}(s, 0, 0, m^2) = \frac{1}{2\p^2},
\end{equation}
proving the validity of a sum rule. 
This demonstrates the exchange of a branch-cut in the on-shell case $(s_1=s_2=0)$ for general massive fermion in the loop rather than poles. The change of the cut into a pole can be illustrated by moving towards the conformal limit. 
 For this purpose we set $m^2_n\equiv (1/n) m^2$ and reduce the value of the mass parameters for $n\to \infty$. \\
The sequence $\Delta^{(n)}{\Phi_0}$, characterising the discontinuity with parameter $m^2_n$, exhibits an interesting behavior, ultimately converging to a Dirac delta function in the limit \(m \to 0\). Specifically, the spectral flow is given by the sequence of functions 
\beq
\lim_{m^2_n\to 0}\Delta^{(n)}{\bar \Phi_0} (s, m_n^2) = \lim_{m_n\to 0} \frac{ m_n^2}{\pi s^2} \log\left(\frac{1 + \sqrt{\tau(s, m_n^2)}}{1 - \sqrt{\tau(s, m_n^2)}}\right) \theta(s - 4m_n^2) = \frac{1}{2\pi} \delta(s),
\eeq
where the convergence to $\delta(s)$ is understood in the distributional sense. The sequence of functions are strongly suppressed at large $s$. This expression encapsulates the essence of how the spectral densities evolve as the mass parameter \(m_n\) approaches zero. 
The pattern is shown in Fig \ref{dd1}. In the figure, the area under the the curve $\Delta^{(n)}{\Phi_0}$ is unrelated to \(m_n\), due to the sum rule \eqref{sumrr} and equals the anomaly. We will refer to this phenomenon as to an "area law" or 
a spectral flow of the sequence of spectral densities $\Delta^{(n)}{\Phi_0}$ towards $s=0$, where the sum rule is indeed saturated by the exchange of an anomaly pole in the conformal limit, while the continuum disappears. We are going to describe the behaviour in the conformal case.

\subsection{On-shell analysis of the conformal anomaly form factor}
At this point, we extend the analysis presented above turning to a description of the on-shell behavior of the $\Phi_{TJJ}$ form factor, that in the presence of a massive fermion loop and pure gauge contributions is given by
\begin{equation}
	\Phi_{TJJ}(s) = \frac{n_F g_s^2 \delta^{ab} \left(3 m^2 (s - 4m^2) \,{C}_0(s,m^2) - 6m^2 + s\right)}{72 \pi^2 s^2} - \frac{11 C_A g_s^2 \delta^{ab}}{144 \pi^2 s},
\end{equation}
where ${C}_0(s,m^2) \equiv C_0(s,0,0,m^2)$ is the standard scalar three-point function (see Appendix \ref{Co}). In the on-shell limit, it takes the form:
\begin{equation}
	{C}_0(s,m^2) = \frac{1}{2s} \log^2\left( \frac{\sqrt{s(s-4m^2)} + 2m^2 - s}{2m^2} \right).
\end{equation}

To study the spectral behavior of $\Phi_{TJJ}$, we compute its discontinuity across the real axis $(s\geq 0)$ using
\begin{equation}\label{lleq1}
	\text{disc}\left(\frac{1}{s}\right)=-2i\pi\delta(s), \qquad \text{disc} \left( \frac{1}{s^2} \right) = 2 i \pi \delta'(s), 
\end{equation}
\begin{equation}\label{lleq2}
	\text{disc} \left( \frac{C_0(s, m^2)}{s} \right)
	= -2i \frac{\pi}{s^2} \log \left( \frac{1 + \sqrt{\tau(s, m^2)}}{1 - \sqrt{\tau(s, m^2)}} \right) \theta(s - 4m^2) - i \pi\,A(0) \delta(s),
\end{equation}

and
\begin{align}\label{lleq4}
	\text{disc} \left( \frac{C_0(s,m^2)}{s^2} \right) = -\frac{2 i \pi}{s^3} \log\left( \frac{1 + \sqrt{\tau(s,m^2)}}{1 - \sqrt{\tau(s,m^2)}} \right) \theta(s - 4m^2) + i \pi \delta'(s) A(s), 
\end{align}
where we defined
\begin{align}
	\tau(s,m^2) = 1 - \frac{4m^2}{s}, 
\end{align}
\begin{align}
	A(s) = {C}_0(s + i\epsilon, m^2) + {C}_0(s - i\epsilon, m^2).
\end{align}
We use the general identity for the distributional difference of powers of propagators gives:
\begin{equation}
	\left( \frac{1}{x + i\epsilon} \right)^n - \left( \frac{1}{x - i\epsilon} \right)^n = (-1)^n \frac{2\pi i}{(n-1)!} \delta^{(n-1)}(x),
\end{equation}
which implies the presence of $\delta(s)$ and $\delta'(s)$ terms in the discontinuities of rational functions like $1/s^n$.

The contribution proportional to $\delta'(s)$ in equation \eqref{lleq4} can be expressed in terms of $\delta$ derivatives acting on $A(s)$:
\begin{equation}
	\delta'(s) A(s) = \delta'(s) A(0) - \delta(s) A'(0), \quad \text{with} \quad A(0) = -\frac{1}{m^2}, \quad A'(0) = -\frac{1}{12 m^4}.
\end{equation}

As a result, the full discontinuity of the fermionic part of $\Phi_{TJJ}$ takes the form
\begin{align}
	\text{disc} \left( \Phi_{TJJ}(s) \right)_{\text{fermion}}
	=\,& \frac{n_F g_s^2 \delta^{ab}}{72 \pi^2} \, \text{disc} \left( \frac{3 m^2 (s - 4m^2) C_0(s,m^2) - 6m^2 + s}{s^2} \right) \\
	=\,& \frac{n_F g_s^2 \delta^{ab}}{72 \pi^2} \biggl[ 3 m^2  \left(\text{disc} \left( \frac{C_0(s,m^2)}{s} \right)+4m^2\text{disc} \left( \frac{C_0(s,m^2)}{s^2} \right)\right) \nonumber\\  &-6m^2  \text{disc} \left( \frac{1}{s^2} \right)  + \text{disc} \left( \frac{1}{s} \right)\biggl]
\end{align}
It is now straightforward to use Eqs.~\eqref{lleq1}, \eqref{lleq2} and \eqref{lleq4} to verify that the $\delta'(s)$ and $\delta(s)$ terms cancel exactly. 
In other words, there is no pole contribution to the spectral density of $\Phi_{TJJ}$ in the fermionic sector, only the continuous cut survives in the on-shell limit. Explicitly, we find
\begin{equation}
	\text{disc}[\Phi_{TJJ}(s)]_{\text{fermion}} = -2i \, \frac{n_F g_s^2 \delta^{ab} \left(3 m^2 (s - 4m^2) \right)}{72 \pi s^3} \log \left( \frac{1 + \sqrt{\tau(s, m^2)}}{1 - \sqrt{\tau(s, m^2)}} \right) \theta(s - 4m^2).
\end{equation}
This result reflects the cancellation of the conformal anomaly pole in the fermionic sector, consistent with the ``area law'' holding for the spectral function.
In contrast, the gluonic contribution to the form factor is given by
\begin{equation}
	\Phi_{TJJ}^{\text{gluon}}(s) = - \frac{11 C_A g_s^2 \delta^{ab}}{144 \pi^2 s},
\end{equation}
whose discontinuity is entirely localized at $s=0$:
\begin{equation}
	\text{disc}[\Phi_{TJJ}(s)]_{\text{pole}} = \frac{11 C_A g_s^2 \delta^{ab}}{144 \pi^2} \cdot 2\pi i \delta(s).
\end{equation}
Therefore, the anomaly pole is absent in the fermionic sector for $m \neq 0$, but remains present in the gluonic sector. In the conformal limit, the branch cut generated by the fermionic contribution over the real axis shrinks to the origin and turns into a pole, while the gluon contributions remain anchored at s=0, being this sector massless. Their combinations restores the full structure of the anomaly pole, whose residue bears the entire conformal anomaly.  The complete anomaly pole re-emerges as
\begin{equation}
	\text{disc}[\Phi_{TJJ}(s)] = - \frac{(2n_F - 11 C_A)\, i g_s^2 \delta^{ab}}{72 \pi} \delta(s).
\end{equation}
Introducing the factor $1/2 i$ coming from the definition of the spectral density we obtain 
\beq
\Delta \Phi_{TJJ}(s)=-\frac{1}{3}\frac{\beta(g)}{g} \delta(s)
\eeq
that reproduces the anomaly sum rule under integration for $s\geq 0$. Notice that the factor $1/3$ comes from 
the definition of the conformal anomaly form factor, due to the transverse projector  ${1}/{ (3 \, q^2)} \hat\pi^{\mu \nu}(q)$ in 
\eqref{anommX}, that is erased once we perform the trace on the correlator. Notice that, in our conventions, the quarks contribute negatively and the gluons positively to the anomaly. 
	
\subsection{Alternative tests using of the Laplace/Borel transforms }
\label{lap}
 We are now going to show that the integral of the spectral density of $\Phi_0$ along the cut from $\left[4 m^2, \infty \right)$ can be 
 evaluated as follows. We use the inverse Laplace transform of the density, denoted as $t$, followed by the $t \to 0$ limit. 
 
 In the most general context, we define the inverse Laplace transform, $\mathcal{L}^{-1}\{F(s)\}$, for a given function $F(s)$ as
\beq
f(t) = \mathcal{L}^{-1}\{F(s)\} = \frac{1}{2\pi i} \int_{\gamma - i \infty}^{\gamma + i \infty} F(s) e^{st} \, ds.
\eeq
Additionally, for a rational function $F(s)$ possessing a pole at $s = s_0$, the inverse Laplace transform is
\beq
\mathcal{L}^{-1}\left\{\frac{1}{s - s_0}\right\} = e^{s_0 t}.
\eeq
This contribution is nonzero as $t \to 0$, making it relevant for the dispersion integral. While this method efficiently computes integrals over the continuous spectrum of the spectral density, caution is required since it does not account for singular contributions, which can appear as subtractions when reconstructing a specific form factor. Notably, double poles, such as $\frac{1}{(s - s_0)^2}$, are not captured in this method, where one obtains
\beq
\mathcal{L}^{-1}\left\{\frac{1}{(s - s_0)^2}\right\} = t e^{s_0 t},
\eeq
which vanishes in the limit $t \to 0$.\\
For the anomaly form factor $\Phi_0$, which has the spectral representation
\beq
\Phi_0(q^2) = \frac{1}{\pi} \int_0^\infty \frac{\Delta \Phi_0(s)}{s - q^2} \, ds,
\eeq
applying the inverse Laplace transform results in
\beq
\mathcal{L}^{-1}\left\{\int_0^\infty \frac{\Delta \Phi_0(s)}{s - q^2} \, ds \right\}(t) = \frac{1}{\pi} \int_0^\infty \Delta \Phi_0(s) e^{st} \, ds.
\eeq
Thus, the integral of the spectral density is evaluated in the $t \to 0$ limit
\beq
\lim_{t \to 0} \mathcal{L}^{-1}\{\Phi_0\}(t) = \frac{1}{\pi} \int_0^\infty \Delta \Phi_0(s) \, ds.
\eeq
This technique provides a means of deriving the integral along the cut and verifying sum rules associated with form factors. 
To apply this method, we utilize the parametric representation of the scalar integrals in $\Phi_0\equiv \Phi_{TJJ}$ from \eqref{tjj} and proceed by direct computation of the inverse Laplace transform. This yields
\small
\begin{equation}
	\mathcal{L}^{-1}\{\Phi_0\}(t) = -\int_0^1 dx \int_0^{1-x} dy \, \frac{g_s^2 \delta^{ab} \left(2 n_f \left(-6 (4 y (x+y-1)+1) e^{-\frac{m^2 t}{y (x+y-1)}} - 30 m^2 t + 24 y (x+y-1) + 7\right) - 11 C_A \right)}{72 \pi^2}.
\end{equation}
\normalsize
This result is valid for the on-shell case, though the extension to more general situations is straightforward, albeit lengthy. Taking the limit as $t \to 0$ leads directly to the value of the integral of the discontinuity across the entire cut
\beq
\mathcal{L}^{-1}\{\Phi_0\}(t \to 0) = \int_{4 m^2}^\infty \Delta \Phi_0(s, p_1^2, p_2^2, m^2) \, ds = \frac{g_s^2}{144 \pi^2}(11 C_A - 2 n_f)=-\frac{1}{3}\frac{\beta(g)}{g}
\eeq
This expression holds in the most general case, including for off-shell vector lines and a massive fermion, and is consistent with the findings in \cite{Giannotti:2008cv}. 
\begin{figure}[t]
\centering
\subfigure[]{\includegraphics[scale=1]{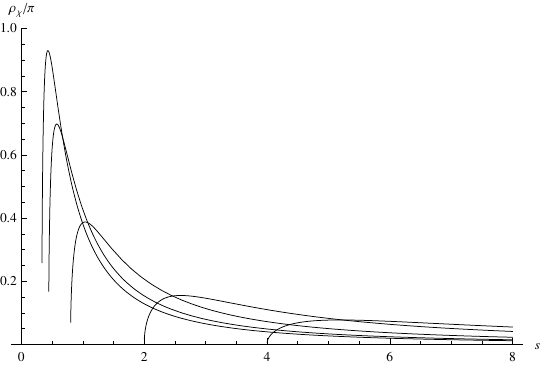}}  \hspace{2cm}
\caption{Chiral spectral density flow in the $p_1^2 = p_2^2 = 0$ case as a function of the fermion mass $m^2_n\equiv m^2/n$ for the $AVV$ for $n\to \infty$. The sum rule is an "area law" for this spectral flow as one moves towards the conformal point  $m\to 0$ and modifies the cuts at $4 m_n^2$ into a pole 
($\sim \delta(s),$ $s\equiv q^2$). The area is subtended by the spectral curves and is preserved by the sequence as $n$ grows and the densities get increasingly peaked at $s=0$. 
In the conformal case, this identical pattern is reproduced only by the fermionic contributions, which are mass-dependent, while the contribution to the anomaly coming from the gluons (not shown) stays always localized at $s=0$, 
proportional to $\delta(s)$. }
\label{dd1}
\end{figure}

\noindent

	\section{The residue of the particle pole  of the conformal anomaly form factor in the light-cone variables }  
	\label{vert}

In order to investigate the possible presence of a massless interpolating state---a particle pole---in the renormalized $TJJ$ interaction (the full vertex), we analyze in this section the residue of the correlator for generic values of the invariants $s_1$ and $s_2$. \\
For the chiral anomaly vertex, the corresponding residue at $q^2=0$ vanishes unless $s_1$, $s_2$, and the fermion mass $m$ are simultaneously set to zero (we adopt the notation $(s_1, s_2) \equiv (p_1^2, p_2^2)$ and $s = q^2 = p_3^2$). This distinctive feature of the interaction~\cite{Coriano:2025ceu} will also appear in the conformal anomaly case. The cancellation arises from delicate interferences among the various tensor components of the correlators. \\
As shown in~\cite{Coriano:2025ceu}, the only tensor structure that survives the limit $s \to 0$ at $s_1 = s_2 = m = 0$ is proportional to the Pontryagin density $F\tilde{F}$ for the chiral gauge anomaly, or to $R\tilde{R}$ in the case of the chiral gravitational anomaly. Thus, when all invariants vanish, the residue reduces to these anomaly-induced structures. This can be interpreted as an \emph{anomaly dominance} of the interaction on the light-cone, since the transverse sectors of the corresponding vertices are subleading. In other words, the light-cone behavior of the interaction is controlled by the anomaly in both cases. A similar result holds in the soft limit, when all four components of $q^\lambda$ are taken to zero. \\
To highlight the difference between the chiral and conformal cases, we now repeat the analysis for the $TJJ$ correlator. Also here it is essential to consider the limit
\begin{equation}
\lim_{s \to 0} \, q^2 \,
\big\langle T^{\mu\nu}(q)\, J^{\alpha a}(p_1)\, J^{\beta b}(p_2) \big\rangle,
\end{equation}
and, for a particle pole to exist, one must show that this expression remains nonzero, regardless of the virtualities of the other invariants. We therefore examine different configurations of the external invariants together with the fermion mass. Two prescriptions for the $q^2 \to 0$ limit will be considered:  
(i) sending the four-momentum $q^\lambda$ along a lightlike direction (light-cone limit), and  
(ii) taking $q^\lambda \to 0$ in all components (soft limit). \\
For the light-cone limit, the four-momenta are parametrized as
\begin{equation}
p_3 \equiv q = q^+ n^+ + q^- n^-, \qquad  
p_1 = p_1^+ n^+ + p_1^- n^- + p_\perp, \qquad  
p_2 = p_2^+ n^+ + p_2^- n^- - p_\perp,
\end{equation}
with light-cone unit vectors satisfying $(n^\pm)^2 = 0$ and $n^+ \cdot n^- = 1$. In this regime, $q^+$, $p_1^+$, and $p_2^+$ are large, while $q^-$, $p_1^-$, and $p_2^-$ are small. The components are given by
\begin{equation}
q^- = \frac{q^2}{2 q^+}, \qquad 
p_1^- = \frac{p_1^2 + p_\perp^2}{2 p_1^+}, \qquad 
p_2^- = \frac{p_2^2 + p_\perp^2}{2 p_2^+}.
\end{equation}
We introduce a scale $\lambda^2 \sim p_1^2 \sim p_2^2$ with $\lambda \ll \bar{Q} = q^+$, where $\bar{Q}$ is large but fixed in the $n^+$ direction. In this scenario, $q_0$ and $q_3$ are large but fixed, while $q^- = q_0 - q_3$ becomes arbitrarily small. The limit $q^2 \to 0$ is then taken at fixed $\bar{Q}$, with $q^- \to 0$, so that any term proportional to $q^+ q^-$ can be neglected. \\
To extract the residue of a tensor structure with open indices, we project onto the light-cone basis. If a component survives this projection, it indicates the presence of a massless $1/q^2$ pole in the residue along the light-cone. The actual behavior of the residue depends on the kinematic configuration of the external momenta and the fermion masses, and is determined by a one-loop computation. \\
For the $AVV$ and $TTJ_5$ correlators, as discussed above, the anomaly form factor contributes with a nonzero residue in the limit $q^2 \to 0$ only when all other invariants vanish. In both cases, the residues match the anomaly coefficients. By contrast, the $TJJ$ correlator behaves differently: the particle pole does not coincide with the anomaly, because an additional pole at zero momentum contributes in the massless, on-shell limit. We will clarify this point in what follows. \\
In particular, if any external momentum is off-shell or if the internal fermion mass is nonzero, the residue of the $TJJ$ correlator vanishes:
\begin{equation}
\lim_{s \to 0} \, q^2 \braket{T^{\mu\nu}(q) J^{\alpha a}(p_1) J^{\beta b}(p_2)} = 0 \qquad  ( \textrm{for } s_1\neq 0 \, \textrm{or}\, s_2\neq 0 \,\textrm{or} \, m\neq 0).
\end{equation}
This mirrors the behavior found in the chiral anomaly case. However, in the conformal case the interaction exhibits a massless pole only in the light-cone configuration
\begin{equation}
\label{mod}
\lim_{s \to 0} \, q^2 \braket{T^{\mu\nu}(q) J^{\alpha a}(p_1) J^{\beta b}(p_2)} \neq 0 \qquad  (  \textrm{for } s_1=0, s_2=0, m= 0).
\end{equation}
The nonzero residue arises upon contracting the correlator with the polarization vectors of the external spin-1 states and imposing transversality. This setup corresponds to a light-cone process, namely a virtual graviton decaying into two on-shell photons or gluons. \\
As discussed above, in the case of the chiral anomalies only the longitudinal tensor structures of the 
$AVV$ and $TTJ_5$ correlators survive, where the longitudinal direction is defined by the momentum of the chiral current,  
showing that the interaction is anomaly dominated. In the case of the $TJJ$, on the other hand, one can verify the validity of 
\eqref{mod}, but two tensor structures survive: one related to the anomaly ($FF$) and a second, traceless one. Specifically, we obtain
\beq\label{tjjpolelim}
\lim_{s \to 0} \, q^2 \braket{T^{\mu\nu}(q) J^{\alpha a}(p_1) J^{\beta b}(p_2)} 
= -\frac{g_s^2 \, \delta^{ab}}{48 \pi^2} \left(\frac{2}{3} n_f - \frac{11}{3} C_A \right) \tilde\phi_1^{\mu\nu\alpha\beta} 
- \frac{g_s^2 \, \delta^{ab}}{288 \pi^2} \left(n_f - C_A \right) \tilde\phi_2^{\mu\nu\alpha\beta}.
\eeq
The analysis in the soft limit is similar, preserving the same structures. 
This result contains two contributions: the first, proportional to $\tilde\phi_1^{\mu\nu\alpha\beta}$, is linked to the conformal anomaly, while the second, associated with $\tilde\phi_2^{\mu\nu\alpha\beta}$, corresponds to an additional pole and is related to a traceless tensor structure. 
The two tensor structures are
\begin{align}
\tilde\phi_1^{\, \mu \nu \alpha \beta} (p_1,p_2) &= 
\left(s \, g^{\mu\nu} - q^{\mu} q^{\nu}\right) \, u^{\alpha \beta} (p_1,p_2),
\label{widetilde1}\\
\tilde\phi_2^{\, \mu \nu \alpha \beta} (p_1,p_2) &= 
- 2 \, u^{\alpha \beta} (p_1,p_2) \left[ s \, g^{\mu \nu} + 2 \left(p_1^\mu p_1^\nu + p_2^\mu p_2^\nu \right)
- 4 \left(p_1^\mu p_2^\nu + p_2^\mu p_1^\nu \right) \right],
\label{widetilde2}
\end{align}
from which it is straightforward to see that $\tilde\phi_2^{\mu\nu\alpha\beta}$ is traceless if the two gluons are taken to be transverse. 
This second traceless structure allows us to identify an additional contribution to the anomaly effective action describing the interaction around the light-cone, of the form
\begin{align}\label{expoleac}
	S_{\text{extra pole}} = \frac{g_s^2}{72 \pi^2} \left(n_f - C_A \right) \int &d^4x \sqrt{-g} \int d^4x' \sqrt{-g'}\, h_{\mu\nu}(x) \Box^{-1}_{x,x'} 
	\nonumber\\&
	\times \left[ 
	3 (\partial^\mu \partial^\nu F^a_{\alpha\beta}) F^{\alpha\beta a}
	+ \frac{1}{4} \left( g^{\mu\nu} \Box - 4 \partial^\mu \partial^\nu \right) F^a_{\alpha\beta} F^{\alpha\beta a}
	\right]_{x'},
\end{align}
while the first tensor structures is associated with the well-known effective action of the form \cite{Armillis:2009pq}
\bea
S_{\text{anom}} &=&\frac{1}{3} \, \frac{g^3}{16 \pi^2} \left (  - \frac{11}{3} \, C_A + \frac{2}{3} \, n_f \right)  \, \int d^4 x \, d^4 y \,R^{(1)}(x)\, \square^{-1}(x,y) \, F^a_{\alpha \beta}F^{\alpha \beta a},
\eea
that generalizes the expression derived in \cite{Giannotti:2008cv} for QED. $R^{(1)}(x)$ is the linearized scalar curvature.
Therefore, in order extract the effective action describing the light-cone behaviour of the correlator, we need to combine the contributions of two functionals generating both $\tilde{\phi_1}$ and 
$\tilde{\phi}_2$ of \eqref{widetilde1} and \eqref{widetilde2}, which are obtained by varying the general light-cone $(\text{lc})$ action 
\beq
\mathcal{S}_{lc}=\mathcal{S}_{\text{anom}} + \mathcal{S}_{\text{extra pole}},
\eeq
 with respect to the external metric fluctuations 
$h^{\mu\nu}$ and non-Abelian gauge fields $A^{\alpha a}$ as 

\begin{equation}
	\left\langle T^{\mu_1\nu_1}(x) J^{\mu_2 a}(y) J^{\mu_3 b}(z) \right\rangle_{\text{lc}}
	= \left. \frac{\delta^3 \mathcal{S}_{\text{lc}}}{\delta h_{\mu_1\nu_1}(x) \delta A^a_{\mu_2}(y) \delta A^b_{\mu_3}(z)} \right|_{h=0,\,A=0}, 
\end{equation}
followed by a Fourier transform to momentum space 
\beq
\Gamma^{\mu_1\nu_1 \mu_2\mu_3 a b }_{lc}=\int d^4 x d^4 y e^{i p_1 x + i p_2 y}\left\langle T^{\mu_1\nu_1}(0) J^{\mu_2 a}(y) J^{\mu_3 b}(z) \right\rangle_{\text{lc}}.
\eeq
The effective action obviously satisfies the trace anomaly constraints 
\bea
g_{\mu\nu}\,  \Gamma_{lc}^{\mu\nu\alpha\beta a b}(p,q) =  - 2 \, \frac{\b (g)}{g}\, u^{\a \b}(p,q) \delta^{a b},
\eea
with 
with the QCD $\b$ function given by \eqref{betagg}. \\
It should be emphasized that this extra contribution does not correspond to an additional effective energy--momentum tensor that is separately conserved, since $\mathcal{S}_{\text{extra pole}}$ is not diffeomorphism invariant.  
Recall that the conservation of the stress--energy tensor is equivalent to the requirement of diffeomorphism invariance of the underlying action.  
In this case, the conservation of the energy--momentum tensor associated with $\mathcal{S}_{\text{extra pole}}$ can be verified only for on-shell vectors, and the absence of diffeomorphism invariance in this action does not contradict the general 
Noether theorem, which holds true for any 
off-shell action which is diffeomorphism invariant.  
This observation aligns with the findings of \cite{Giannotti:2008cv}, which analyzed similar phenomena within the framework of QED. The specific tensor structure associated with $\mathcal{S}_{\text{extra pole}}$ excludes the interpretation of the amplitude as arising from scalar particle exchange. As observed in \cite{Giannotti:2008cv}, the emergence of a pole featuring a nontrivial tensor character can be traced to non-vanishing contributions from terms such as $\partial^\mu \partial^\nu F_{\alpha\beta}$ and $\partial^\mu \partial^\nu (F_{\alpha\beta}F^{\alpha\beta})$, evaluated in the presence of external gauge field backgrounds. These background configurations inherently violate Lorentz symmetry, and the resulting massless excitations 
are not scalar in nature.  \\
In summary, our analysis shows that, in restricted light-cone kinematics, the residue at the particle pole of the \(TJJ\) correlator differs from its \(AVV\) and \(TTJ_5\) counterparts~\cite{Coriano:2025ceu}. In general it is not solely tied to the trace anomaly; only in the special case \(n_f = N_c = C_A\) does \(\mathcal{S}_{\text{extra pole}}\) vanish. This suggests that the chiral and gravitational anomalies impose stricter constraints on the relation between the particle-pole residue and light-cone behaviour, making the anomaly--correlator interplay more intricate than in the other two cases.

\section{Comments: the local effective actions for axions and dilatons}
The use of a local action to describe an asymptotic state raises significant questions, particularly regarding the treatment of axion and dilaton interactions at the phenomenological level. Specifically, if an anomaly is associated with an anomaly pole, and this pole does not necessarily represent an asymptotic state (i.e., a particle pole), one might argue that a local action transforming a pole into an asymptotic state has limited applicability.\\
Typically, the local interaction is derived by coupling the divergence of an anomaly current— in this case, the dilaton current— to the dilaton field, thus generating the ordinary local $\rho FF$ interaction. Interactions of this form, such as the dilaton case or the $\alpha(x) F\tilde{F}$ form in the axion case, are naturally obtained by applying Ward identities to the underlying anomaly vertex. However, when moving away from the conformal limit, at least in a perturbative framework, these interactions appear not to be consistently motivated.\\
In the context of confining gauge theories, this issue can be effectively resolved by recognizing the breakdown of the perturbative picture when describing certain anomaly interactions below the nonperturbative confinement scale. For example, in the case of axions, the local action and its interactions are derived through manipulations of the chiral fermion measure in the path integral. These manipulations are equivalent to applying Ward identities, and the off-shell effective action derived from perturbative expansion takes a local form.\\
The framework we have explored is closely tied to the conformal properties of such vertices, which are clearly broken by nonperturbative effects. A second possibility to endorse a local effective description emerges from the analysis of similar interactions in the presence of a Higgs phase. Both the axion and dilaton cases are expected to be consistent with the results derived in the UV through the analysis of all such anomalies, supported by anomaly matching arguments in both the UV and the IR, only in these two cases.

\section{Conclusions}
We have shown that sum rules associated with interactions linked to the conformal anomaly impose crucial constraints on the absorptive parts of the corresponding anomaly form factor. Our findings reveal that, under a wide range of kinematic conditions, the spectral density of this anomaly can exhibit unexpected behaviors, including the emergence of dilaton-like interpolating states in the correlators. Depending on the kinematic configuration, contributions may cancel, or appear as poles or cuts in the complex plane of the stress-energy tensor's momentum. Notably, even when mass effects convert a pole into a cut in the spectral densities, the sum rule remains effectively satisfied due to the emergence of a pole. This behavior reflects the analytic structure of the amplitudes and the intrinsic characteristics of the interactions.\\
The breaking of conformal symmetry is thus intimately tied to the appearance of massless intermediate states. Combining this analysis with our earlier study of chiral and gravitational anomaly form factors, we observe that in the conformal and on-shell limits, anomaly poles behave as true propagating degrees of freedom. Conversely, the introduction of mass effects and renormalization corrections complicates the spectral structure but preserves the sum rule. These phenomena have been explored through both perturbative methods and momentum-space conformal field theory ($CFT_p$), which decomposes correlation functions into transverse and longitudinal components.\\
It would be interesting to investigate how the specific constraints on the partonic correlator, as we have explored, manifest at the hadronic level, particularly in the analysis of gravitational form factors for the pion and proton. Another noteworthy aspect of this analysis is its potential implications for general gravitational scatterings in the eikonal limit involving photons and gravitons, which will be explored in forthcoming works.
\vspace{0.3cm}
\centerline{\bf Acknowledgements}
This work is partially supported by INFN, inziativa specifica {\em QG-sky}, by the the grant PRIN 2022BP52A MUR "The Holographic Universe for all Lambdas" Lecce-Naples, and by the European Union, Next Generation EU, PNRR project "National Centre for HPC, Big Data and Quantum Computing", project code CN00000013.

\appendix

\section{Relevant functions}\label{appfunc}
\subsection{Quark Trace Sector}
The quark massive part in the trace of the correlator (anomaly free) is given by
\begin{equation}
		g_{\m\n}\braket{T^{\m\n}J^{\a a}J^{\b b}}_q= +B_1(p_1\cdot p_2) g^{\a\b}+ B_2 p_1^{\a}p_1^{\b}+B_2(p_1\leftrightarrow p_2) p_2^{\a}p_2^{\b}+B_3 p_1^{\a}p_2^{\b}+B_4p_1^{\b}p_2^{\a}
\end{equation}
where the coefficient $B_i$ with $i=1,2,3,4$ are given by
{\smaller\allowdisplaybreaks
	\begin{align}
		B_1=& - \frac{g_s^2}{16\p^2} m^2 n_f   \Biggl(\frac{3     \left(p_1^6-p_1^4 \left(p_2^2+3 q^2\right)-p_1^2 \left(p_2^2-3 q^2\right) \left(p_2^2+q^2\right)+\left(p_2^2-q^2\right)^3\right) C_0\left(p_1^2,p_2^2,q^2,m^2\right)}{p_1^4-2 p_1^2 \left(p_2^2+q^2\right)+\left(p_2^2-q^2\right)^2}
		\nn\\
		&\qquad \qquad \qquad +\frac{6   p_1^2 \bar B_0(p_1^2,m^2) \left(p_1^2-p_2^2-q^2\right)}{p_1^4-2 p_1^2 \left(p_2^2+q^2\right)+\left(p_2^2-q^2\right)^2} 
		+\frac{6 q^2\bar  B_0(q^2,m^2)   \left(q^2 \left(p_1^2+p_2^2\right)-\left(p_1^2-p_2^2\right)^2\right)}{q^2 \left(p_1^4-2 p_1^2 \left(p_2^2+q^2\right)+\left(p_2^2-q^2\right)^2\right)} \nn\\
		&\qquad \qquad \qquad  -\frac{6 p_2^2
			\bar B_0(p_2^2,m^2)\left(p_1^2-p_2^2+q^2\right)}{p_1^4-2 p_1^2 \left(p_2^2+q^2\right)+\left(p_2^2-q^2\right)^2} +6 \Biggl)-12   m^4 \frac{g_s^2}{16\p^2} C_0\left(p_1^2,p_2^2,q^2,m^2\right)
	\end{align}

	\begin{align}
		B_2= & - \frac{g_s^2}{16\p^2} m^2 n_f \Biggl(\frac{12    p_2^2  \left(p_1^6-p_1^4 \left(p_2^2+q^2\right)-p_1^2 \left(p_2^4-6 p_2^2 q^2+q^4\right)+\left(p_2^2-q^2\right)^2 \left(p_2^2+q^2\right)\right) C_0\left(p_1^2,p_2^2,q^2,m^2\right)}{\left(p_1^4-2 p_1^2 \left(p_2^2+q^2\right)+\left(p_2^2-q^2\right)^2\right)^2}\nn\\ &\qquad \qquad \qquad+\frac{24    p_2^2 p_1^2\bar B_0(p_1^2,m^2)\left(2 p_1^4-p_1^2 \left(p_2^2+q^2\right)-\left(p_2^2-q^2\right)^2\right)}{p_1^2 \left(p_1^4-2 p_1^2 \left(p_2^2+q^2\right)+\left(p_2^2-q^2\right)^2\right)^2}\nn\\ &\qquad \qquad \qquad-\frac{24    p_2^2q^2\bar B_0(q^2,m^2)\left(q^2 \left(p_1^2+p_2^2\right)+\left(p_1^2-p_2^2\right)^2-2 q^4\right)}{q^2 \left(p_1^4-2 p_1^2 \left(p_2^2+q^2\right)+\left(p_2^2-q^2\right)^2\right)^2}\nn\\ &\qquad \qquad \qquad-\frac{24    p_2^2\bar B_0(p_2^2,m^2) \left(p_1^4+p_1^2 \left(p_2^2-2 q^2\right)-2 p_2^4+p_2^2 q^2+q^4\right)}{\left(p_1^4-2 p_1^2 \left(p_2^2+q^2\right)+\left(p_2^2-q^2\right)^2\right)^2}\nn\\ &\qquad \qquad \qquad+\frac{24    p_2^2  }{p_1^4-2 p_1^2 \left(p_2^2+q^2\right)+\left(p_2^2-q^2\right)^2}\Biggl)-\frac{3   m^4 p_2^2 g_s^2 C_0\left(p_1^2,p_2^2,q^2,m^2\right)}{\p^2\left(p_1^4-2 p_1^2 \left(p_2^2+q^2\right)+\left(p_2^2-q^2\right)^2\right)}
	\end{align}
	
	\begin{align}
		B_3=&- \frac{g_s^2}{16\p^2} m^2 n_f\Biggl(\frac{6   \left(p_1^2+p_2^2-q^2\right) \left(p_1^6-p_1^4 \left(p_2^2+q^2\right)-p_1^2 \left(p_2^4-6 p_2^2 q^2+q^4\right)+\left(p_2^2-q^2\right)^2 \left(p_2^2+q^2\right)\right) C_0\left(p_1^2,p_2^2,q^2,m^2\right)}{\left(p_1^4-2 p_1^2 \left(p_2^2+q^2\right)+\left(p_2^2-q^2\right)^2\right)^2} \nn\\ &\qquad \qquad \qquad+\frac{12  p_1^2\bar B_0(p_1^2,m^2)\left(p_1^2+p_2^2-q^2\right) \left(2 p_1^4-p_1^2 \left(p_2^2+q^2\right)-\left(p_2^2-q^2\right)^2\right)}{p_1^2 \left(p_1^4-2 p_1^2 \left(p_2^2+q^2\right)+\left(p_2^2-q^2\right)^2\right)^2}\nn\\ &\qquad \qquad \qquad-\frac{12   q^2\bar B_0(q^2,m^2) \left(-3 q^4 \left(p_1^2+p_2^2\right)+4 p_1^2 p_2^2 q^2+\left(p_1^2-p_2^2\right)^2 \left(p_1^2+p_2^2\right)+2 q^6\right)}{q^2 \left(p_1^4-2 p_1^2 \left(p_2^2+q^2\right)+\left(p_2^2-q^2\right)^2\right)^2}\nn\\ &\qquad \qquad \qquad-\frac{12  p_2^2\bar B_0(p_2^2,m^2) \left(p_1^2+p_2^2-q^2\right) \left(p_1^4+p_1^2 \left(p_2^2-2 q^2\right)-2 p_2^4+p_2^2 q^2+q^4\right)}{p_2^2 \left(p_1^4-2 p_1^2 \left(p_2^2+q^2\right)+\left(p_2^2-q^2\right)^2\right)^2}\nn\\ &\qquad \qquad \qquad+\frac{12    \left(p_1^2+p_2^2-q^2\right)}{p_1^4-2 p_1^2 \left(p_2^2+q^2\right)+\left(p_2^2-q^2\right)^2}\Biggl)-\frac{3   m^4 g_s^2  \left(p_1^2+p_2^2\right) C_0\left(p_1^2,p_2^2,q^2,m^2\right)}{2\p^2\left(p_1^4-2 p_1^2 \left(p_2^2+q^2\right)+\left(p_2^2-q^2\right)^2\right)}
	\end{align}
	
	\begin{align}
		B_4=&- \frac{g_s^2}{16\p^2} m^2 n_f \, p_1\cdot p_2 \Biggl(\frac{6   \left(p_1^2+p_2^2-q^2\right) \left(p_1^6-p_1^4 \left(p_2^2+3 q^2\right)+p_1^2 \left(-p_2^4+10 p_2^2 q^2+3 q^4\right)+\left(p_2^2-q^2\right)^3\right) C_0\left(p_1^2,p_2^2,q^2,m^2\right)}{\left(p_1^4-2 p_1^2 \left(p_2^2+q^2\right)+\left(p_2^2-q^2\right)^2\right)^2}\nn\\ &\qquad \qquad \qquad\qquad+\frac{12  p_2^2\bar B_0(p_2^2m^2) \left(-5 p_1^4+4 p_1^2 \left(p_2^2+q^2\right)+\left(p_2^2-q^2\right)^2\right)}{\left(p_1^4-2 p_1^2 \left(p_2^2+q^2\right)+\left(p_2^2-q^2\right)^2\right)^2}\nn\\ &\qquad \qquad \qquad\qquad+\frac{12   p_1^2\bar B_0(p_1^2,m^2)\left(p_1^4+p_1^2 \left(4 p_2^2-2 q^2\right)-5 p_2^4+4 p_2^2 q^2+q^4\right)}{\left(p_1^4-2 p_1^2 \left(p_2^2+q^2\right)+\left(p_2^2-q^2\right)^2\right)^2}\nn\\ &\qquad \qquad \qquad\qquad-\frac{12   q^2\bar B_0(q^2,m^2)\left(q^4 \left(p_1^2+p_2^2\right)+\left(p_1^2-p_2^2\right)^2 \left(p_1^2+p_2^2\right)-2 q^2 \left(p_1^4-4 p_1^2 p_2^2+p_2^4\right)\right)}{q^2 \left(p_1^4-2 p_1^2 \left(p_2^2+q^2\right)+\left(p_2^2-q^2\right)^2\right)^2}\nn\\ &\qquad \qquad \qquad\qquad+\frac{12   \left(p_1^2+p_2^2-q^2\right)}{p_1^4-2 p_1^2 \left(p_2^2+q^2\right)+\left(p_2^2-q^2\right)^2}\Biggl)-  \frac{ 3\, m^4 \,  g_s^2 \left(p_1^2+p_2^2-q^2\right)^2 C_0\left(p_1^2,p_2^2,q^2,m^2\right)}{2\left(p_1^4-2 p_1^2 \left(p_2^2+q^2\right)+\left(p_2^2-q^2\right)^2\right)}
	\end{align}}

\begin{align}
		\chi_0=&g_s^2m^4 H_1 C_0 \left(p_1^2,p_2^2,q^2,m^2\right) +g_s^2m^2 \Bigl(H_2C_0 \left(p_1^2,p_2^2,q^2,m^2\right)  \nn\\&+ H_3 \bar{B}_0(q^2,m^2)+ H_4 \bar{B}_0(p_1^2,m^2)+ H_5 \bar{B}_0(p_2^2,m^2) +H_6\Bigl)
\end{align}
with $H_i$ with $i=1,2,3,4$ given by
\begin{equation}
	H_1= n_f\frac{3 \left(p_1^4+p_2^4+q^4-2 \left(p_1^2+p_2^2\right) q^2\right)  }{2 \pi ^2 \left(-p_1^2-p_2^2+q^2\right) \left(q^4-2 \left(p_1^2+p_2^2\right) q^2+\left(p_1^2-p_2^2\right)^2\right)}
\end{equation}
\begin{equation}
	H_2=- n_f\frac{3 \tilde P_1}{8 \pi ^2 \left(-p_1^2-p_2^2+q^2\right) \left(q^4-2 \left(p_1^2+p_2^2\right) q^2+\left(p_1^2-p_2^2\right)^2\right)^2}
\end{equation}
\begin{equation}
	H_3= n_f\frac{3 \, \tilde P_2 }{4 \pi ^2 q^2 \left(-p_1^2-p_2^2+q^2\right) \left(q^4-2 \left(p_1^2+p_2^2\right) q^2+\left(p_1^2-p_2^2\right)^2\right)^2}
\end{equation}
\begin{equation}
	H_4=-n_f\frac{3\, p_1^2\left(-3 p_2^6+\left(p_1^2+5 q^2\right) p_2^4+\left(p_1^4-q^4\right) p_2^2-\left(q^2-p_1^2\right)^3\right) }{4 \pi ^2 \left(p_1^2+p_2^2-q^2\right) \left(q^4-2 \left(p_1^2+p_2^2\right) q^2+\left(p_1^2-p_2^2\right)^2\right)^2} 
\end{equation}
\begin{equation}
	H_5= n_f\frac{3\, p_2^2\left(p_2^6+\left(p_1^2-3 q^2\right) p_2^4+\left(p_1^4+3 q^4\right) p_2^2-\left(q^2-p_1^2\right)^2 \left(3 p_1^2+q^2\right)\right)}{4 \pi ^2 \left(-p_1^2-p_2^2+q^2\right) \left(q^4-2 \left(p_1^2+p_2^2\right) q^2+\left(p_1^2-p_2^2\right)^2\right)^2} 
\end{equation}
\begin{equation}
	H_6=n_f \frac{3 \left(p_1^4+p_2^4+q^4-2 \left(p_1^2+p_2^2\right) q^2\right) }{4 \pi ^2 \left(-p_1^2-p_2^2+q^2\right) \left(q^4-2 \left(p_1^2+p_2^2\right) q^2+\left(p_1^2-p_2^2\right)^2\right)} 
\end{equation}

with
\begin{align}
	\tilde P_1= &-p_2^{10}+\left(p_1^2+5 q^2\right) p_2^8-2 q^2 \left(3 p_1^2+5 q^2\right) p_2^6+2 \left(5 q^2-3 p_1^2\right) q^2 \left(p_1^2+q^2\right) p_2^4\nonumber\\&-\left(q^2-p_1^2\right)^2 \left(p_1^2+q^2\right) \left(5 q^2-p_1^2\right) p_2^2+\left(q^2-p_1^2\right)^5,
\end{align}

\begin{align}
	\tilde{P}_2=&q^2 \Bigl(-p_2^8+\left(2 p_1^2+3 q^2\right) p_2^6-\left(p_1^2+q^2\right) \left(2 p_1^2+3 q^2\right) p_2^4\nn\\&+\left(q^2-p_1^2\right) \left(-2 p_1^4+3 q^2 p_1^2+q^4\right) p_2^2+p_1^2 \left(q^2-p_1^2\right)^3\Bigl).
\end{align}

 \section{The scalar off-shell, massive 3-point function} 
\label{Co}
\begin{equation}
\label{lco}
		C_0\left(q^2,p_1^2,p_2^2,m^2\right)= \frac{1}{i \pi^2}\int d^n k\frac{1}{(k^2- m^2)( (k- q)^2- m^2) 
		((k - p_1)^2- m^2)},\,
	\end{equation}
($s_1=p_2^2, s_2=p_2^2$) 
\allowdisplaybreaks{
\begin{align}
C_0(s,s_1,s_2,m^2)=&\frac{\text{Li}_2\left(\frac{s_1 \sqrt{\lambda(s,s_1,s_2)}-s_1 (-s+s_1-s_2)}{s_1 (s-s_1+s_2)+\sqrt{s_1 \left(s_1-4 m^2\right)} \sqrt{\lambda(s,s_1,s_2)}} -\left(i \left(-s+s_1-s_2-\sqrt{\lambda(s,s_1,s_2)}\right)\right) \epsilon \right) }{ \sqrt{\lambda(s,s_1,s_2)}}\nonumber\\&
+\frac{\text{Li}_2\left(\frac{(s+s_1-s_2) s_2+\sqrt{\lambda(s,s_1,s_2)} s_2}{(s+s_1-s_2) s_2+\sqrt{s_2 \left(s_2-4 m^2\right)} \sqrt{\lambda(s,s_1,s_2)}} -\left(i \left(-s-s_1+s_2-\sqrt{\lambda(s,s_1,s_2)}\right)\right) \epsilon \right) }{ \sqrt{\lambda(s,s_1,s_2)}}\nonumber\\&
+\frac{\text{Li}_2\left(\frac{s \sqrt{\lambda(s,s_1,s_2)}-s (s-s_1-s_2)}{ -s (s-s_1-s_2)-\sqrt{s \left(s-4 m^2\right)} \sqrt{\lambda(s,s_1,s_2)} } -\left(i \left(-s+s_1+s_2+\sqrt{\lambda(s,s_1,s_2)}\right)\right) \epsilon \right)}{ \sqrt{\lambda(s,s_1,s_2)}}\nonumber\\&
+\frac{\text{Li}_2\left(\frac{s \sqrt{\lambda(s,s_1,s_2)}-s (s-s_1-s_2)}{\sqrt{s \left(s-4 m^2\right)} \sqrt{\lambda(s,s_1,s_2)}-s (s-s_1-s_2)}+\epsilon  \left(i \left(s-s_1-s_2-\sqrt{\lambda(s,s_1,s_2)}\right)\right)\right) }{\sqrt{\lambda(s,s_1,s_2)}}\nonumber\\&
+\frac{ \text{Li}_2\left(\frac{(s+s_1-s_2) s_2+\sqrt{\lambda(s,s_1,s_2)} s_2}{(s+s_1-s_2) s_2-\sqrt{s_2 \left(s_2-4 m^2\right)} \sqrt{\lambda(s,s_1,s_2)}}+\epsilon  \left(i \left(s+s_1-s_2+\sqrt{\lambda(s,s_1,s_2)}\right)\right)\right)}{\sqrt{\lambda(s,s_1,s_2)}}\nonumber\\&
+\frac{\text{Li}_2\left(\frac{s_1 \sqrt{\lambda(s,s_1,s_2)}-s_1 (-s+s_1-s_2)}{s_1 (s-s_1+s_2)-\sqrt{s_1 \left(s_1-4 m^2\right)} \sqrt{\lambda(s,s_1,s_2)}}+\epsilon  \left(i \left(s-s_1+s_2+\sqrt{\lambda(s,s_1,s_2)}\right)\right)\right) }{\sqrt{\lambda(s,s_1,s_2)}}\nonumber\\&
-\frac{\text{Li}_2 \left(\frac{-s (s-s_1-s_2)-\sqrt{\lambda(s,s_1,s_2)} s}{-s (s-s_1-s_2)-\sqrt{s \left(s-4 m^2\right)} \sqrt{\lambda(s,s_1,s_2)}} -\left(i \left(-s+s_1+s_2-\sqrt{\lambda(s,s_1,s_2)}\right)\right) \epsilon \right)}{\sqrt{\lambda(s,s_1,s_2)}}\nonumber\\&
-\frac{\text{Li}_2 \left(\frac{-s_1 (-s+s_1-s_2)-\sqrt{\lambda(s,s_1,s_2)} s_1}{s_1 (s-s_1+s_2)+\sqrt{s_1 \left(s_1-4 m^2\right)} \sqrt{\lambda(s,s_1,s_2)}} -\left(i \left(-s+s_1-s_2+\sqrt{\lambda(s,s_1,s_2)}\right)\right) \epsilon \right)}{ \sqrt{\lambda(s,s_1,s_2)}}\nonumber\\&
-\frac{\text{Li}_2 \left(\frac{(s+s_1-s_2) s_2-s_2 \sqrt{\lambda(s,s_1,s_2)}}{(s+s_1-s_2) s_2+\sqrt{s_2 \left(s_2-4 m^2\right)} \sqrt{\lambda(s,s_1,s_2)}}- \left(i \left(-s-s_1+s_2+\sqrt{\lambda(s,s_1,s_2)}\right)\right) \epsilon \right)}{ \sqrt{\lambda(s,s_1,s_2)}}\nonumber\\&
-\frac{\text{Li}_2 \left(\frac{(s+s_1-s_2) s_2-s_2 \sqrt{\lambda(s,s_1,s_2)}}{(s+s_1-s_2) s_2-\sqrt{s_2 \left(s_2-4 m^2\right)} \sqrt{\lambda(s,s_1,s_2)}}+\epsilon  \left(i \left(s+s_1-s_2-\sqrt{\lambda(s,s_1,s_2)}\right)\right)\right)}{\sqrt{\lambda(s,s_1,s_2)}}\nonumber\\&
-\frac{\text{Li}_2 \left(\frac{-s_1 (-s+s_1-s_2)-\sqrt{\lambda(s,s_1,s_2)} s_1}{s_1 (s-s_1+s_2)-\sqrt{s_1 \left(s_1-4 m^2\right)} \sqrt{\lambda(s,s_1,s_2)}}+\epsilon  \left(i \left(s-s_1+s_2-\sqrt{\lambda(s,s_1,s_2)}\right)\right)\right)}{\sqrt{\lambda(s,s_1,s_2)}}\nonumber\\&
-\frac{\text{Li}_2 \left(\frac{-s (s-s_1-s_2)-\sqrt{\lambda(s,s_1,s_2)} s}{\sqrt{s \left(s-4 m^2\right)} \sqrt{\lambda(s,s_1,s_2)}-s (s-s_1-s_2)}+\epsilon  \left(i \left(s-s_1-s_2+\sqrt{\lambda(s,s_1,s_2)}\right)\right)\right)}{\sqrt{\lambda(s,s_1,s_2)}}.
\label{cco}
\end{align}
}
In \eqref{cco} we have used the short-hand notation $i \varepsilon f(s,s_1,s_2)$ as $ i \varepsilon \textrm{ sign} f(s,s_1,s_2) $, 
with $\textrm{ sign}$ the signum function. $\lambda(s,s_1,s_2)$ is the K\"allen function
\begin{equation}
	\lambda(s,s_1,s_2)=s^2-2 s_1 s-2 s_2 s+s_1^2+s_2^2-2 s_1 s_2
\end{equation}
factorizable as
\begin{equation}
	\lambda(s,s_1,s_2)=(s-(\sqrt{s_1}-\sqrt{s_2})^2)(s-(\sqrt{s_1}+\sqrt{s_2})^2).
\end{equation}
\subsection{Massless limits of $C_0$ and of renormalized $B_0$}
 For $m=0$ 
\begin{equation}
\label{lco}
		C_0\left(q^2,p_1^2,p_2^2\right)= \frac{1}{i \pi^2}\int d^n k\frac{1}{k^2 (k- q)^2 (k - p_1)^2},\,
	\end{equation}
 is the the ordinary massless scalar master integral, with
 
\bea
 &  C_0 ( p_1^2,p_2^2,q^2) = \frac{ 1}{q^2} \Phi (x,y),
 \eea
where the function $\Phi (x, y)$  is defined as
\bea
\Phi( x, y) &=& \frac{1}{\lambda} \biggl\{ 2 [Li_2(-\rho  x) + Li_2(- \rho y)]  +
\ln \frac{y}{ x}\ln \frac{1+ \rho y }{1 + \rho x}+ \ln (\rho x) \ln (\rho  y) + \frac{\pi^2}{3} \biggr\},
\label{Phi}
\eea
with
\bea
 \lambda(x,y) = \sqrt {\Delta},
 \qquad  \qquad \Delta=(1-  x- y)^2 - 4  x  y,
\label{lambda} \\
\rho( x,y) = 2 (1-  x-  y+\lambda)^{-1},
  \qquad  \qquad x=\frac{p_1^2}{q^2} \, ,\qquad \qquad y= \frac {p_2^2}{q^2},\, 
\eea

\begin{equation}
{B}_0(p^2)=\sdfrac{1}{i\p^\frac{d}{2}}\int\,d^d l\ \frac{1}{l^2(l-p)^2}=\frac{ \left[\G\left(\frac{d}{2}-1\right)\right]^2\G\left(2-\frac{d}{2}\right)}{\G\left(d-2\right)(p^2)^{2-\frac{d}{2}}}\label{B0ex}.
\end{equation}
We define $(\epsilon=d-4)$
\beq
B_0(p^2)=\frac{1}{\varepsilon}+\bar{B}_0(p^2)
\eeq
and the renormalized expression of $B_0(p^2,0)$ used in the spectral analysis is 
\beq
B_0^R(p^2,0)\equiv \bar{B}_0(p^2)= 2 + \log(\mu^2/p^2) 
\label{BB}
\eeq
in the $\overline{MS}$ scheme, with $\mu$  the renormalization scale. \\

\subsection{ $C_0$ at $s_{\pm}=(\sqrt{s_1}\pm \sqrt{s_2})^2$}
The evaluation of $C_0$ at the points $s_\pm$ are given by
\begin{align}
\label{x1}
	C_0(s_-,s_1,s_2,m^2)=&\frac{1}{{s_1}{s_2}\, s_-^2}\Biggl(\sqrt{{s_1}} \Biggl(\sqrt{{s_2} \left({s_2}-4 m^2\right)} \left(\sqrt{{s_1}}-\sqrt{{s_2}}\right)\times \nn\\ &
	\times \log \left(\frac{\sqrt{{s_2} \left({s_2}-4 m^2\right)}+2 m^2-{s_2}}{2 m^2}\right)\nonumber\\&+\sqrt{{s_2}} \sqrt{\left(-2 \sqrt{{s_1}} \sqrt{{s_2}}+{s_1}+{s_2}\right) \left(-4 m^2-2 \sqrt{{s_1}} \sqrt{{s_2}}+{s_1}+{s_2}\right)} \log \left(S_1\right)\Biggl)\nonumber\\&+\sqrt{{s_2}} \sqrt{{s_1} \left({s_1}-4 m^2\right)} \left(\sqrt{{s_2}}-\sqrt{{s_1}}\right) \log \left(\frac{\sqrt{{s_1} \left({s_1}-4 m^2\right)}+2 m^2-{s_1}}{2 m^2}\right)\Biggl)
\end{align}
and 
\begin{align}
\label{x2}
	C_0( s_+,s_1,s_2,m^2)=&\frac{1}{{s_1}{s_2} s_+^2}\Biggl(\sqrt{{s_1}} \Biggl(\sqrt{{s_2} \left({s_2}-4 m^2\right)} \left(\sqrt{{s_1}}+\sqrt{{s_2}}\right)\times \nn \\&
	\times \log \left(\frac{\sqrt{{s_2} \left({s_2}-4 m^2\right)}+2 m^2-{s_2}}{2 m^2}\right)\nonumber\\&-\sqrt{{s_2}} \sqrt{\left(\sqrt{{s_1}}+\sqrt{{s_2}}\right)^4-4 m^2 \left(\sqrt{{s_1}}+\sqrt{{s_2}}\right)^2} \log \left(S_2\right)\Biggl)\nonumber\\&+\sqrt{{s_2}} \sqrt{{s_1} \left({s_1}-4 m^2\right)} \left(\sqrt{{s_1}}+\sqrt{{s_2}}\right) \log \left(\frac{\sqrt{{s_1} \left({s_1}-4 m^2\right)}+2 m^2-{s_1}}{2 m^2}\right)\Biggl), 
\end{align}
where
\begin{equation}
	S_1=\frac{\sqrt{\left(-2 \sqrt{{s_1}} \sqrt{{s_2}}+{s_1}+{s_2}\right) \left(-4 m^2-2 \sqrt{{s_1}} \sqrt{{s_2}}+{s_1}+{s_2}\right)}+2 m^2-\left(\sqrt{{s_1}}-\sqrt{{s_2}}\right)^2}{2 m^2}
\end{equation}
and 
\begin{equation}
	S_2=\frac{\sqrt{\left(\sqrt{{s_1}}+\sqrt{{s_2}}\right)^4-4 m^2 \left(\sqrt{{s_1}}+\sqrt{{s_2}}\right)^2}+2 m^2-\left(\sqrt{{s_1}}+\sqrt{{s_2}}\right)^2}{2 m^2}.
\end{equation}

\subsection{Gluon Trace Sector}
\label{gg}
The coefficient functions are given by
{\allowdisplaybreaks
	\begin{eqnarray}
		C_1 &=& -\frac{ \,  \,  \, C_A \, g_s^2 \,  }{ 32\pi^2 \big( (p_1 \cdot p_2)^2 - p_1^2 \, p_2^2 \big)} \, \Big(
		2 \, (p_1 \cdot p_2)^2 - 2 \, p_1^2 \, p_2^2 
		- p_1^2 \, (p_1 \cdot p_2 + p_2^2 ) \bar B_0 (p_1^2)  \notag \\ &&
		+ \big[ p_2^2 \, (p_1^2 - p_1 \cdot p_2) - 2 \, (p_1 \cdot p_2)^2 \big]   \bar B_0(p_2^2)
		+ (p_1 \cdot p_2) \, (2 \, p_1 \cdot p_2 + p_1^2 + p_2^2) \,  \bar B_0 (q^2)  \notag \\ &&
		+ \big[ 2\, (p_1 \cdot p_2+p_1^2) (p_1 \cdot p_2)^2\,  - p_2^2 \, \big(4 (p_1 \cdot p_2)^2   + p_1^4\big) + 5 \, p_1^2 \, p_4^4  \big] C_0(p_1^2, p_2^2 , q^2)  \Big) \\
		C_2 &=& -2  \,  \,  \, C_A \, \frac{g_s^2}{16\pi^2} \,  \, (p_1 \cdot p_2) \, C_0 (p_1^2 , p_2^2 , q^2) \\
		C_3 &=& -\frac{ \,  \,  \, g_s^2 \, C_A \, (p_1^2 + p_2^2) \,}{32\pi^2 \big(p_1^2 \, p_2^2 - (p_1 \cdot p_2)^2 \big)} \Big( 
		(-p_1\cdot p_2-p_1^2)   \bar B_0(p_1^2)+(-(p_1\cdot p_2)-p_2^2) \,  \bar B_0(p_2^2) \notag \\ &&
		+q^2 \, \bar  B_0(q^2)+\big(p_1^2 \, (p_1\cdot p_2-2 p_2^2) +(p_1\cdot p_2) \, \big(4 \, (p_1\cdot p_2)+p_2^2) \big) \, C_0(p_1^2,p_2^2, q^2)  \Big)  \\
		C_4 &=& \, \,  \, \frac{g_s^2}{32\pi^2} \, C_A \,  \Big( (p_1^2 - p_2^2) \, \big[  \bar B_0 (p_1^2) - \bar  B_0(p_2^2) \big] 
		+ \big[ p_1^4 + p_2^4 - 2 \, (p_1^2 + p_2^2) \, p_1 \cdot p_2 - 6 \, p_1^2 \, p_2^2 \big] \, C_0(p_1^2, p_2^2, q^2) \Big). \nonumber \\
\end{eqnarray}}

\providecommand{\href}[2]{#2}\begingroup\raggedright\endgroup


\end{document}